\documentclass[a4paper,11pt]{article}
\usepackage{pos}

\usepackage{adjustbox}

\usepackage{adjustbox}
\begin{document}

\title{Domain Walls and Hubble Constant Tension}
\ShortTitle{Tension, Domain Walls}                    

\author*[a]{Holger Bech Nielsen}
\author[b]{Colin D. Froggatt}

\affiliation[a]{Niels Bohr Institute,   \\
  Jagtvejen 155a, Copenhagen, Denmark}

\affiliation[b]{School of Physics and Astronomy, Glasgow University,
Glasgow G12 8SU, Great Britain}


\emailAdd{hbech@nbi.dk}
\emailAdd{Colin.Froggatt@glasgow.ac.uk}


\FullConference{Corfu Summer Institute 2023 "School and Workshops on Elementary Particle Physics and Gravity" (CORFU2023)\\
 23 April - 6 May , and 27 August - 1 October, 2023\\
Corfu, Greece\\}

\abstract{We present the idea that replacing the cosmological constant
  $\Lambda$ in the
  $\Lambda CDM$ model by a distribution of walls, with very low tension compared
  to what one would expect from 
  ``new physics'', could help explaining the tension in the Hubble
  constant fits in the Standard Cosmological Model. Using parameters from our
  model for dark matter as macroscopic pearls, we can get a
  promising order of magnitude for the correction to the Hubble constant estimated from observations of the cosmic microwave background.
  Our model is on the borderline of failing by predicting too large
  extra fluctuations as a function of  direction in the cosmological microwave
  background
  radiation. However, imagining the bubbles in the voids to have come from
  more somewhat smaller ``big bubbles'' also occurring outside the big voids
  may help. We estimate that, in order to have big volumes of the new vacuum
in intergalactic space, a very high temperature is needed and that such regions
would be likely to get cooled, ``freeze'' and shrink down to the degenerate
form of dark matter if hitting some ordinary matter, as is likely in the
denser parts of the Universe.
We also review our model for dark matter, and develop the understanding of the
stopping of the dark matter particles in the shielding of say the DAMA-LIBRA
underground experiment and the counting rate this experiment observes. We manage to obtain a consistent fit with a mass $M=2*10^{-18}\, kg= 10^{9}\, GeV$ and radius $R=10^{-10}m$ for the dark matter pearls, corresponding to a
tension in the domain wall of $S=(8\, MeV)^3$.}

\maketitle
\section{Introduction}
Both the Hubble LeMaitre constant $H_0$ tension and the existence of  dark matter
seem to call
for some effectively new physics, while the most remarkable result from accelerators is that they do {\em not} provide any surviving signs of
new physics. The Hubble constant tension consists of the disagreement between local measurements giving $H_0 = 73.8 \pm 1.0$ km/s/Mpc \cite{Local} and the value $H_0 = 67.4 \pm 0.5$ km/s/Mpc \cite{CMB} based on cosmic microwave background (CMB) radiation measurements by the Planck satellite.

We have for some time written about our speculative model of dark matter\cite{Dark1,Dark2, Tunguska, Corfu17, Corfu19, supernova, extension,Corfu21,Bled20,Bled21,Bled22,Bled23,BS,theline} in which
the new physics consists of the assumption that there should be several - at least 2 - vacuum phases,
with very much the same energy density\cite{Dvali, MPP1,MPP2,Corfu1995,MPP3,MPP4,Picek}. For the moment we
speculate that the
difference between the two phases of vacuum could be due to some details in the
way the pion fields and the sigma field in the Nambu-Jonalasinio model form
the vacua, when the quark masses are not exactly negligible. If this is true,
even the several phases of vacuum would not be truly new physics in the
sense of needing an explicit violation of the Standard Model; it would
just be that one had not exactly calculated that there can be more phases of
the strong interaction vacuum than previously thought. In any case it would
be nice if also the Hubble-LeMaiatre constant tension could be explained from
the same several vacua story. In fact the idea of the present article is to
study the possibility that regions of what we may call vacuum 2 (a vacuum
phase different from the one we live in) could deliver an explanation for this
tension, by adding to the energy of the Universe and especially some excess energy to the microwave
background radiation. Indeed that is what is very likely
in our model, because we have already assumed for a long time - in order to keep some
ordinary matter inside the bubbles of vacuum 2 under high pressure so as
to form heavily concentrated dark matter pearls - that there is a potential
difference for nucleons inside vacuum 2 versus  the vacuum in which we live.
This potential difference, called $\Delta V \simeq 3$ MeV, is directed so as to pull the
nucleons into vacuum 2. Such energy, of course not considered in models without
our several vacua, would cause a ``kinetic energy production''
by nucleons falling into vacuum 2, so that it can, as we shall see, modify the
energy going into the cosmic microwave background radiation, so as to perturb
the estimated Hubble-LeMaitre expansion rate.

So our main subject is:

 {\em Resolution of the Tension in the Hubble Constant fitting in the Standard
    Cosmological Model by different Degenerate Vacua
  with a surprisingly small
    Domain Wall Energy Density (Tension) $S$.}

    \subsection{Plan of Paper}
    
    In the following section \ref{arguments} we shall give arguments supporting our  main
    hypothesis that there are several phases of the vacuum, at least two for
    the purpose of the present article. Next in section \ref{tension}  we
    briefly mention the famous problem that the Standard Cosmological Model
    get a ca 9\% inconsistency for the value of the Hubble-Lemaitre constant \cite{Local,CMB},
    and we give the main idea by which we wish to solve this problem, namely
    that an appreciable amount of  nuclei get caught by the ``vacuum 2''
    in which they have a lower potential. As a consequence they ``release'' energy,
    some of which ends up increasing the temperature $T$ of the microwave
    background radiation, so that a higher value has been measured than
    what would have appeared in the Standard Model not having our second
    vacuum phase. In section \ref{fitting} we then calculate what is needed and
    estimate that our model with the suggested potential difference for a
    nucleon in the two different phases $\Delta V \simeq 3$ MeV, under the assumption
    of huge and comparable extensions of both vacua, has a good chance to
    fit well. Part of this estimation is contained in section \ref{DeltaT}.

    Next in section \ref{stopping} we discuss another problem for our model, which is important for the underground dark matter experiments, namely how strongly our
    dark matter pearls get stopped when passing through the earth shielding
    above such experiments as e.g. DAMA-LIBRA. In section \ref{homolumo} we
    shortly review how we consider the X-ray observations of a barely
    observable otherwise hard to fit line at $3.5$ keV as telling us that
    the interior matter in the vacuum 2 of the dark matter pearls should have
    a homolumo gap equal to $3.5$ keV and thus crudely a density
    $\rho_B=5*10^{11}kg/m^3$.

    In section \ref{DAMAlike} we review the somewhat mysterious situation
    for underground dark matter searches, that only {\em one} experiment
    sees the dark matter, namely DMA-LIBRA \cite{DAMA1,DAMA2}, while all the others, such as ANAIS \cite{ANAIS} and Xenon experiments \cite{Xe1T,XenT} do not
    see any dark matter in open conflict with
    DAMA-LIBRA. Our main explanation is that our pearls have stopping lengths in earth
    so that they first essentially stop at the somewhat deeper place of DAMA-LIBRA than the
    other experiments such as ANAIS, and that the pearls cannot stop
    at all in the experimental apparatuses based on fluid xenon. One
    should understand that in our model we take it that the signal DAMA-LIBRA
    sees is a radiation, presumably of electrons, coming ``slowly'' from
    the dark matter pearls, so that it may require a little time, in
    which the dark matter should stop or move slowly to be observed.
    The counting rate of events in the DAMA-LIBRA experiment is considered in section \ref{solution}.
    In section \ref{table} we then collect up the information of our fitting
    to mainly the DAMA-LIBRA observations for the size of the pearls, and
    find that indeed there is a possibility for a consistent solution,
    which we present in Table \ref{t}.

    In section \ref{cosmology} we take up the question as to whether it is possible for
    the domain wall tension $S=(8\, MeV)^3$ fitted to the DAMA-Libra results etc. to allow
    the very  extended regions of vacuum 2 needed for resolving the
    Hubble-Lemaitre constant tension. The attempts to get consistency
    point in the direction that it would be best for the cosmology to have
    a lower tension $S$ for the domain wall between the vacua than the
    $(8\, MeV)^3$ suggested by our fitting to DAMA-LIBRA etc.
We shall argue in section
\ref{hot} that, in order to have large regions of vacuum 2 with our
expected parameters, it is necessary to have very hot plasma in the
vacuum 2 bubbles. If they get cooled down, they will not be able to keep
up the pressure from the surrounding domain wall and they will shrink.
They therefore have the best chance to survive without shrinking away
out in the big voids, where there is almost no matter at all; then
nothing can cool them down except the microwave background radiation, and
that cooling is extremely inefficient if the plasma in the vacuum 2 regions
have densities like in the voids and temperatures of the order of MeV, as the
nucleons immediately get when
they are driven into the voids by the potential difference $\Delta V$.
    However the temperature needed to
    keep even a very large bubble of vacuum 2 from collapsing in outer space
    is embarrassingly high. Finally in section \ref{behind} we discuss the second order phase  transition curve found from lattice simulations of the QCD vacuum in the so-called ''Columbia Plot", as a function of the quark masses. We speculate that the physical quark masses lie on this curve and that the phase transition would connect our two vacua. 

    In section \ref{conclusion} we review and conclude the article.

    \section{Arguments for Several Vacuum Phases (with same energy density)}
    \label{arguments}
%
  \begin{itemize}
  \item There is a 4$\sigma$ evidence for the {\em fine structure constant
    $\alpha$ varying} by about $\frac{\Delta \alpha}{\alpha}\sim 10^{-5}$
    between different places in the Universe \cite{svfsc,svfsc2}.

    It is of course the idea, that if there are several different types or
    phases of vacuum, then they would presumably have different values
    of the natural constants such as e.g. the fine structure constant. Thus
    clouds of gas or plasma lying in different vacuum phase regions would show
    different values of the fine structure constant $\alpha$. In fact using
    our estimate of the different potential energy in the two phases supposed
    to be the extensive ones in space, we got the suggestion \cite{walls} that the deviation
    was of the relative order of $10^{-5}$, which fits very well with the
    observed spatial variation of the fine structure constant.  It must though
    be admitted that popular fitting of the spatial variation of the fine
    structure constant by a dipole variation is not so favourable for our model.
    In fact a two-phase model for the spatial variation of the fine structure
    constant would suggest that $\alpha$ would only take on {\em two}
    values, rather than varying in a continuous manner as the dipole model
    suggests. The dipole model seems favoured, but the uncertainties are so
    big that no variation at all is still very possible.
    
  \item We have had success with a model for {\em dark matter as
    small pieces of a new vacuum phase filled with ordinary matter under
    pressure.} \cite{Dark1,
    Dark2, Tunguska, Corfu17, Corfu19, supernova, extension,Corfu21, Bled20,Bled21,Bled22,Bled23,BS,theline}.
    The major need for the second vacuum is to provide a domain wall separating
    the two vacua with a sufficiently high tension $S$ that it can keep
    some highly compressed ordinary matter inside a bubble of vacuum 2 - surrounded then of course by the domain wall. Its specific weight becomes
    so exceedingly high, $\rho_B=5*10^{11}kg/m^3$, that the dark matter  pearl thereby formed becomes so small compared to its mass, that even the big mass of dark matter is not
    seen significantly. Our dark matter pearls may not be totally invisible,
    say nearby, but compared to their relatively huge mass they are essentially invisible.

  \item Some lattice simulations of the QCD-vacuum with varying temperature indicate
    a {\em phase transition} as a function of the {\em quark masses}
    \cite{question,Columbiaplot,Japanese,Guenther}.
    In section \ref{behind} below, we shall refer to some
    computer-simulations indicating that there is a phase transition
    for the vacuum as a function of the light quark masses, u,d,s.
    If this phase transition is true and Nature realizes it (by having
    chosen just appropriate quark masses), then our story of a second vacuum
    would not be truly ``new physics'', because it would just be a matter
    of using the Standard Model and calculating that there are several phases
    just in the Standard Model.
    
  \item We previously {\em pre}dicted the Higgs mass before the Higgs boson was
    found in LHC {\em from this principle of several vacua with the same energy
      density} \cite{tophiggs}. That is to say, that in the time in the 1990's
    when it was believed that the Higgs boson existed, but  had not
    yet been seen, we found that one could obtain
   a second vacuum in the Standard Model  with the same energy density
    (= cosmological constant) as the vacuum we live in by adjusting the 
    Higgs mass parameter, and we found that by this
    adjustment you would get a Higgs mass of 135 Gev $\pm$ 10 GeV. {\em Later}
    - in 2012 - the Higgs mass was found to be 125 GeV.
  \item {\em Several Vacua may replace the need for the mysterious concept of 
    Dark Energy}
    \begin{figure}
      \includegraphics[scale=0.4]{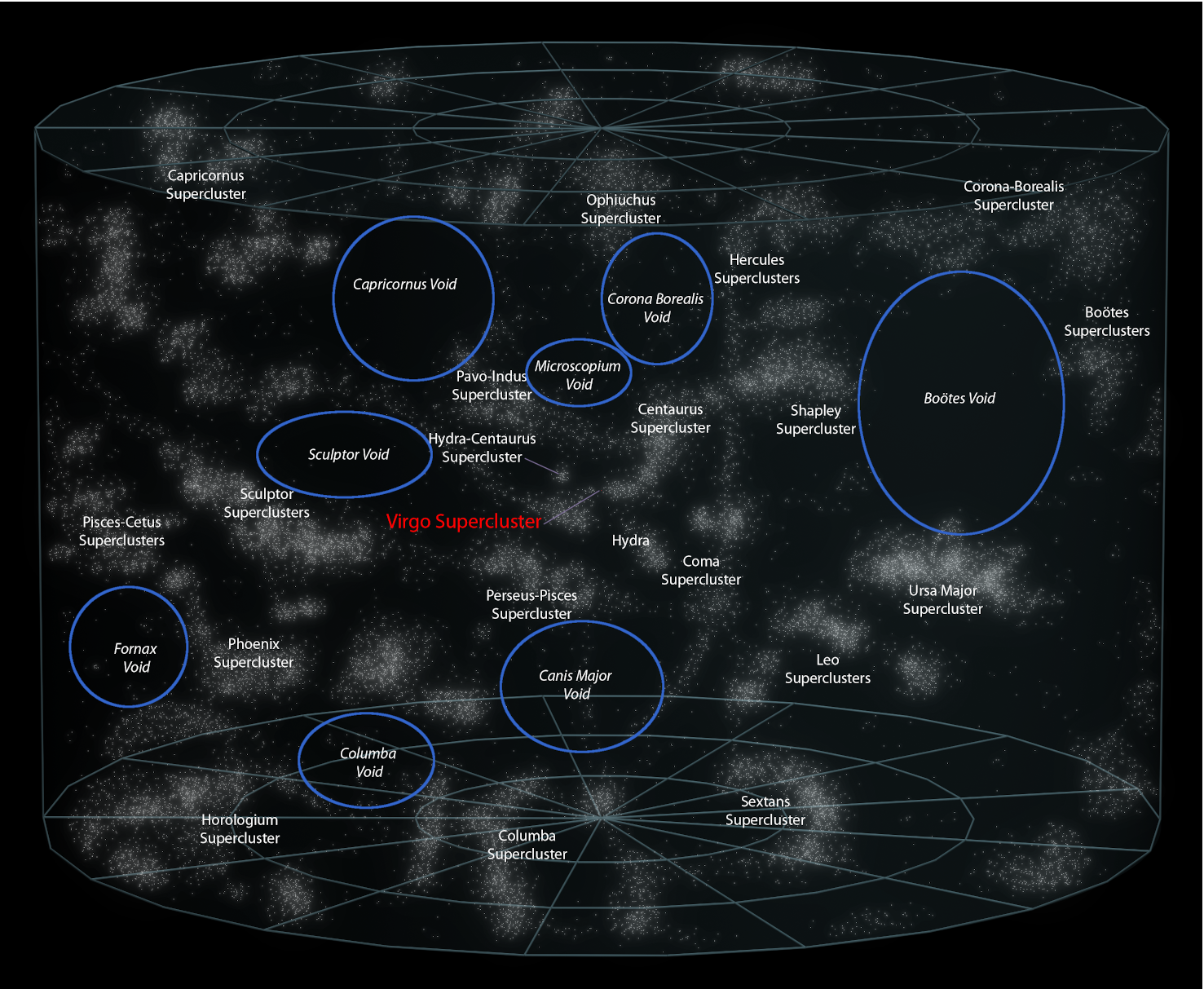}
      \caption{\label{voidsfig}Here we see a drawing in perspective of the
        galaxy distribution in the Universe, and one sees the huge voids
        encircled by ellipses. One hypothesis we have much liked was that
        the vacuum 2 regions were mainly lying in these voids. One could, if so,
        consider the encircling ellipses as representing the domain walls around
        the big vacuum 2 regions filling most of the void in question.
        Even if the vacuum 2 regions  are very large and kept away from
        galaxies and matter, it would be hard for them to keep sufficiently
        hot to be able to keep the domain walls spanned out (see section \ref{hot}). The big vacuum 2
        bubbles would be made to collapse or contract by any interaction with
        some ordinary matter, because compared to the bubbles it would be too
        cold.}
    \end{figure}
    
    Our idea that astronomical voids (big regions in space with rather few galaxies) correspond to big regions of vacuum 2 surrounded by domain walls \cite{walls} is illustrated in Figure \ref{voidsfig}.
    The domain walls have a negative pressure similar to that of 
    dark energy, and can - only effectively by a factor of 2/3 - {\em replace the cosmological constant} \cite{wallreplacing}.

    If one could replace the dark energy by the similar negative pressure in the
    astronomically extended domain walls, they might
    re-contract and open up the possibility that the Universe
    would not expand forever. It would of course then still be a mystery
    why the cosmological constant would be so close to zero (or truly zero),
    so that problem would not be solved.

  \item Generically the coefficient $\theta$ in the Lagrange term $\theta
    F^{\mu\nu}F^{\rho\sigma}\epsilon_{\mu\nu\rho\sigma}$ will be effectively
    different in the different vacua and thus the extension of such
    vacua could adjust to minimize the energy in much the same way as
    the axion field is meant to adjust. Thus the several vacua may {\em replace
    the axion theory.} (See \cite{StrongCPMasao} for an earlier idea).

    One should have in mind that the $\theta$-term is only topological in
    the sense that the action is an integral only depending on the boundary
    conditions and instantons; but now if there were a somewhat different
    (effective) $\theta$ in the different vacua, then the adjustment of the
    domain wall could adjust the effective $\theta$ for the whole  world
    (the average $\theta$). This is very similar to what the axion field
    does. It can adjust the effective $\theta$, so as to explain why it is
    so surprisingly small. So we say: with the different phases, we do not
    need the axion any more.
    \end{itemize}
  \section{The Hubble-Lemaitre Constant Tension}
  \label{tension}
  \begin{itemize}
  \item {\em Our main idea for the solution to the Hubble constant tension problem is that energy is released by
    nuclei passing from our
    vacuum to the ``new'' vacuum 2.}
    
    We should have in mind, that there is a lower potential for the nuclei
    in the vacuum 2 than in the vacuum we live in. Thus, when nuclei move from
    the vacuum we live in into the vacuum 2, energy is released which is of course not counted in models without the extra vacuum.

    In order to make the idea more concrete let us imagine that, in the early universe after the big bang, 
   some relatively few bubbles of vacuum 2 expand by
    taking in more and more nuclei - which in turn may drive more
    electrons into the vacuum 2 region - and that some bubbles of this type
    manage to become so large that they can keep spanned out, in spite of
    the domain wall seeking to contract due to its tension. Much
    of the energy released by the passage of the nuclei into the bubble
    will presumably heat up the plasma in the bubble.

  \item This released energy finds its way by the Sunyaev-Zeldovich (SZ) mechanism \cite{SZ}
    mainly to the cosmic microwave background (CMB) radiation and brings
    the temperature up from an original one (say 2.4 K) to the
    presently observed 2.725 K.

    Inside a big bubble with very low plasma density, the nuclei and
    electrons will only very seldomly hit each other and then have the chance
    to radiate light. So a way to lose a significant amount of energy would be
    for the particles in the plasma to interact with the microwave background
    radiation and deliver heat to this radiation. This is what is called the
    Sunyaev-Zeldovick-mechanism and is the way in which some of the
    extra energy comes to increase the temperature of the CMB. So the
    temperature without this effect of our vacuum 2 bubbles would have been
    smaller than what we measure. 
    
  \item A lower temperature, as we should have gotten without this SZ-effect
    from nuclei passing through the walls, would
    correspond to a larger red shift
    $z= z_{r}$ for the (re)combination of of atoms (usually taken to be
    370000 years after the big bang) than the usual value $z_{r} \simeq   1100$, and
    this may {\em explain the tension.}

    The usual redshift $z_{r}$ at the time when the atoms formed is essentially
    the ratio of the temperature of plasma recombining so that it becomes
    transparent relative to the temperature of the CMB today. But if we now
    invent a mechanism of some ``extra'' energy heating up the CMB-temperature
    today, then we are measuring a temperature misleading us to believe that
    the Universe has expanded less than what it really has expanded since
    the recombination time. This could of course lead to a bad fitting of the
    cosmology.
    
    \end{itemize}
  \subsection{Our Model, also for Dark Matter}

  Our suggestion of there existing two - or more (there is indeed more
  since above we had a third type of vacuum involved in our prediction of the
  Higgs mass) - vacuum phases is applied in a couple of different ways
  in our attempt to resolve mysteries in physics and astronomy:

  \begin{itemize}
  \item We have the big bubbles of vacuum 2 presumably extended to
    such big regions as the big voids observed between the clusters and filaments
    of galaxies. These should be imagined as having very thin and very hot  plasma
    inside, because otherwise they would upset the average density
    of matter in the universe, which we must consider essentially known.
    Or else they would have to be much smaller than the big voids.

  \item Secondly we study much smaller atomic size or nano-meter size
    bubbles, that should make up the dark matter. The smaller a bubble is
    the higher the pressure needed to keep such a bubble from collapsing,
    since the pressure needed is
    \begin{eqnarray}
      P&=& \frac{2S}{R},
      \end{eqnarray}
    where $R$ is the radius of the bubble and $S$ the tension in the domain
    wall.

    In the small pearls supposed to make up the dark matter, the enormous
    pressure is supposed to come about by the internal ordinary matter being so
    compressed that its density is highly increased, and the electrons are
    degenerate as in an ordinary metal. We do though believe that looking
    at it in more detail it is not exactly  a metal, but that it  has a gap
    - the homolumo gap - between the empty and the filled electron energy levels
    of $3.5$ keV, because we want to make electrons crossing this gap
    give the $3.5$ keV X-ray radiation line observed astronomically from some galaxy clusters etc. This X-ray radiation is believed to originate from dark matter.
    But the pressure needed to keep the pearls from collapsing implies a high density, which means that the pearls are very small and have very little visibility
    compared to their mass. So they are essentially dark, having very little
    interaction per mass with light. This property provides a good model for dark matter.

    \end{itemize}

\begin{figure}[h]  
	\includegraphics[scale=0.75]{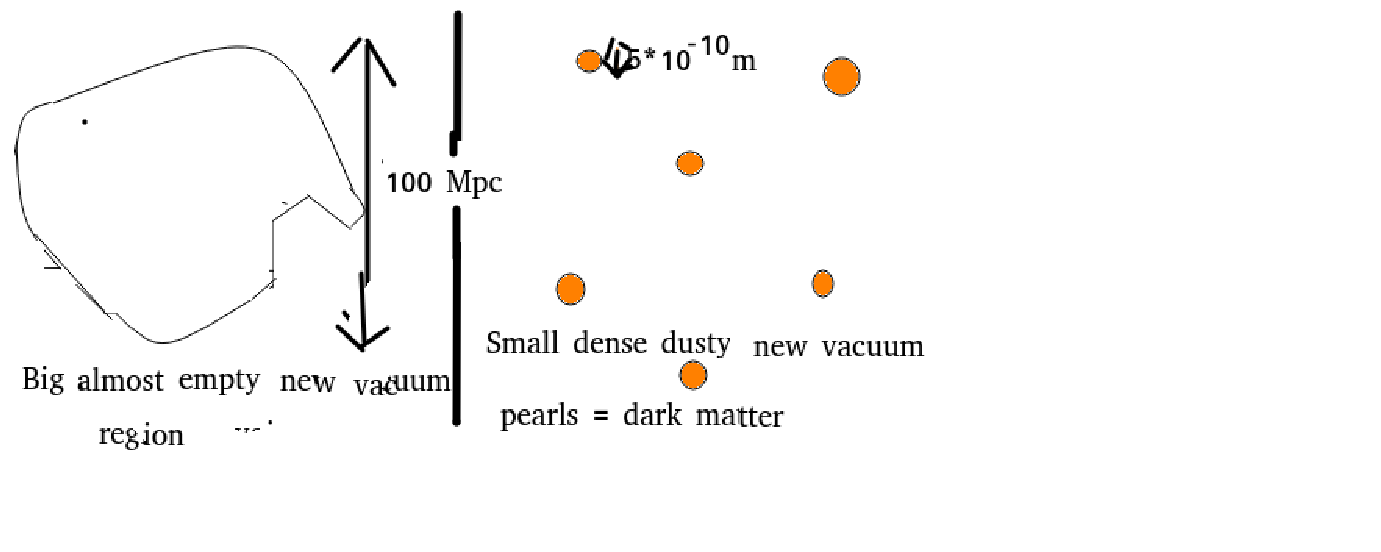}
	\caption{\label{bigsmall}
		Here we illustrate the two kinds of new vacuum regions with a big
		bubble on the left and some small bubbles to the right. The big bubbles must
		be hot to avoid contracting themselves - or it might help lowering the
		needed temperature to have them united into fewer percolating regions - while the small ones keep themselves spanned out by the pressure of the	degenerate electrons}
\end{figure}  

 {\em Two kinds of New Vacuum Regions: Big and Small}

\begin{figure}[h]
  	\includegraphics[scale=0.75]{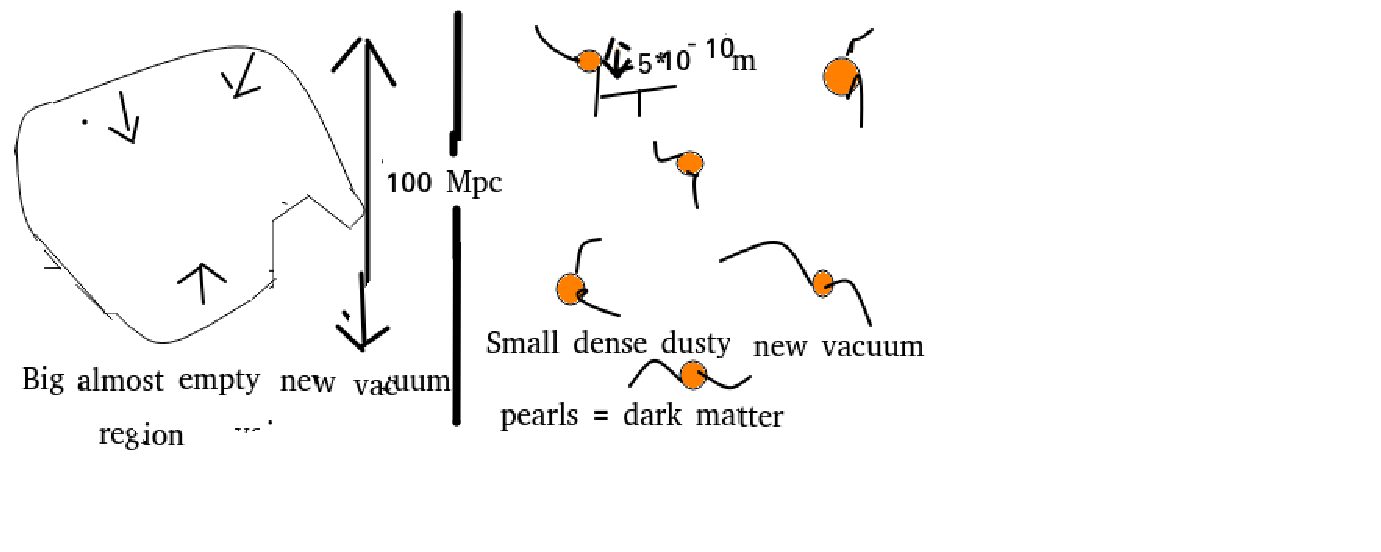}
  	\caption{\label{smalldusty}
  		In this figure  the  small bubbles have been
  		provided with the dust as small black strings attached to them. The
  		big bubble has got some arrows pointing into the bubble
  		illustrating how the nuclei will run into the bubble driven by the potential
  		difference and thus produce heat, which in turn is what causes the
  		Hubble-Lemaitre constant tension.}
  \end{figure}

  In Figures \ref{bigsmall} and \ref{smalldusty} we illustrate the two types of bubble containing the new vacuum. The small bubbles are extremely dense with a radius of order $10^{-10}\, m$, collect dust and are identified with dark matter. The big bubbles with a size of order 100 Mpc are identified with the voids and nuclei are attracted into them with energies of several MeV. 
  
  \begin{figure}[h]
  	\includegraphics[scale=0.8]{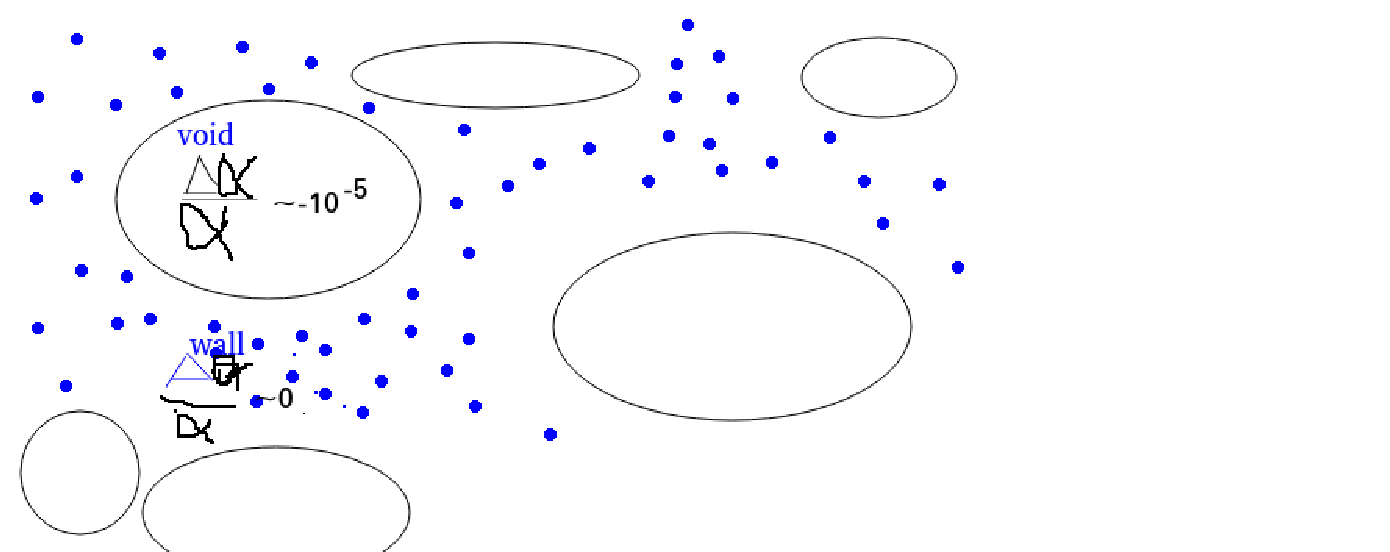}
  	\caption{\label{finestructure} 
  		''Artistic impression of Vacua in the Universe2; Voids contain vacuum 2, clusters of galaxies contain “our” vacuum.
  		The small dots symbolise galaxies of course. The formulas should illustrate that we imagine that the fine
  		structure constant $\alpha$ should deviate a little bit $\frac{\Delta\alpha}{\alpha}$ from the usual value, or say the value on Earth, when one
  		measures it in a void. The formula suggests that the deviation is relatively of the order of $10^{-5}$
  		in the voids,}
  \end{figure}
 
 In Figure \ref{finestructure} we illustrate how the big bubbles can have
 fine structure constant values deviating slightly from the standard value
 and are mainly placed in the big voids.

  \section{Fitting of Sound Horizon $r_s$ and Angular Diameter Distance $d_A$}
  \label{fitting}


	
We use the following CMB fit parameters in standard notation:
\begin{eqnarray}
	\Omega_b h^2 &=& 0.02237 \label{Ob}\\
	\Omega_m h^2 &=& 0.143 \\
	\Omega_{\Lambda} &=& 0.685\\
	h &=& 0.674 \label{CMBh}\\
	T &=& 2.7255 K\\
	\Omega_{\gamma} h^2 &=& 2.47*10^{-5}\label{Og}\\
	\Omega_{R} h^2 &=& 4.15*10^{-5}\\
	\frac{\Omega_b}{\Omega_{\gamma}} &=& 906  \label{bgammaratio}\\
	T &=& 2.7255K \\
	z_r	&=& 1100
\end{eqnarray}
In this fit to the microwave background data $h$ is the Hubble-Lemaitre
constant relative to the value $100km/s/Mpc$, and the $\Omega$'s are the
energy  densities for the various types of
matter contributing to the energy density expressed as a fraction of the critical energy density	:
\begin{itemize}
\item{$\Omega_b$} baryons,
\item{$\Omega_m$} dark plus baryonic matter,
\item{$\Omega_{\Lambda}$} Vacuum energy density or dark energy,
\item{$\Omega_{\gamma}$}  light, cosmic microwave background radiation,
\item{$\Omega_R$}  Total radiation meaning photons and neutrinos, which we take to be massless. 
\end{itemize}

The temperature of the microwave background today is denoted $T$, and
$z_r$ is the redshift for the time of the recombination of the atoms, which for illustration we take to be $z_r = 1100$.

  {\em Calculation}:

	We calculate $r_s$ and $d_A$ using the following expressions 
	\begin{eqnarray}
		r_s &=& 3000 Mpc\int_{z_r}^{\infty}\frac{dz}{H(z)}c(z)\\
		d_A &=& 3000 Mpc\int_{1}^{z_r}\frac{dz}{H(z)}\\
		H(z) &=& \sqrt{\Omega_mh^2(1+z)^3 + \Omega_Rh^2(1+z)^4 +
                  \Omega_{\Lambda}h^2}\\
		c(z)^2 &=& \frac{1}{3(1+\frac{3\Omega_b}{4\Omega_{\gamma}(1+z)})}
	\end{eqnarray}

Using the listed parameters we obtain
\begin{eqnarray}
	r_s &=& 143.86 Mpc\\
	d_A &=& 13949 Mpc
\end{eqnarray}	


Now, motivated by our model outlined in section \ref{tension}, we consider {\em reducing the temperature and increasing the Hubble
  constant}
to the locally measured value {\em h = 0.738}, keeping the "physical" mass
densities (measured in units of
$\sim 1.9*10^{-26} kg/m^3$) $\Omega_m h^2$ and $\Omega_bh^2$ fixed. We
note $\Omega_{\Lambda}$ is negligible in $r_s$ and $\Omega_{R}$ is negligible
in $d_A$. It follows that $r_s$ is essentially unchanged by the change in $h$ and that $d_A$ is essentially unaffected by the change in temperature.

  
We list below the results for three different temperatures and give the
percentage changes $\Delta r_s$  and $\Delta d_A$ in $r_s$ and $d_A$.
 \begin{itemize}
 	\item T = 2.5 K $z_r = 1200$ h = 0.738
 	
 	$r_s$ = 142.69 Mpc \quad $\Delta r_s = -0.81\%$
 	
 	$d_A$ = 13721 Mpc \quad $\Delta d_A = -1.7\%$  
 	
 		\item T = 2.4 K $z_r = 1250$ h = 0.738
 	
 	$r_s$ = 141.75 Mpc \quad $\Delta r_s = -1.5\%$
 	
 	$d_A$ = 13730 Mpc \quad $\Delta d_A = -1.6\%$  
 	
 		\item T = 2.3 K $z_r = 1300$ h = 0.738
 	
 	$r_s$ = 140.00 Mpc \quad $\Delta r_s = -2.7\%$
 	
 	$d_A$ = 13739 Mpc \quad $\Delta d_A = -1.5\%$  
 \end{itemize}
 
 So the precisely measured (with an accuracy of 0.03$\%$) angular size of the acoustic horizon at recombination
 $\theta_{\star} = \frac{r_s}{d_A}$ is {\em kept fixed} if
 the drop in
 temperature to {\em around T= 2.4 K} is compensated by an increase in the
 Hubble
 constant to equal the larger locally  measured value, provided we assume that
 $\Omega_m h^2$ and $\Omega_bh^2$ are kept fixed.

  {\em In order to resolve the Hubble constant tension in our model, we need the CMB temperature unmodified by the SZ mechanism \cite{SZ} to be $T=2.4K$.}  Of course we included in our calculation, that a change in the temperature prior
  to the SZ-heating up meant a change in the early $\Omega_R$. 
  
 It is not clear what effect
 this change in $T$ and $h$ would have on the other observables in the
 complicated CMB fit.

 \section{Change in Temperature $\Delta T$}
 \label{DeltaT}
We now consider how the temperature change of order
\begin{equation}
\Delta T = 2.73\,K - 2.4\,K = 0.33\,K	\label{deltaT}
\end{equation}
might arise in our model.
As briefly discussed in section \ref{homolumo} for our small dark matter pearls we obtain a potential difference $\Delta V$ for nucleons between the two vacuum phases of the order
\begin{equation}
	\Delta V = 3 \, \hbox{MeV}.
\end{equation}
This value $\Delta V =$ 3 MeV was estimated \cite{theline} by requiring the material inside the pearl to have a gap - a homolumo gap - between the empty and the filled electron states of 3.5 keV, so that the dark matter preferably emits X-rays with the observed energy of the mysterious X-ray emission line reported \cite{Bulbul,Boyarsky,Boyarsky2} in 2014.  

  So now let us make a very crude estimate of the order of magnitude of the shift in the observed CMB temperature due to the SZ effect, which we have argued would be good for solving the Hubble-Lemaitre tension:  T = 2.725 K $\rightarrow$  T = 2.4 K. The effect should be most active in the
  last few milliards of years, because effects from earlier times would tend
  to be washed out by being Hubble-Lemaitre expanded away.
  We shall use the ratio
  \begin{eqnarray}
    \frac{\Omega_{b}}{\Omega_{\gamma}} &=& 906
    \end{eqnarray}
    from equation \ref{bgammaratio} above,
  where $\Omega_b$ is the baryon density relative to the critical density,
  and $\Omega_{\gamma}$ the relative density of light, meaning the CMB. Then we make the following assumptions:

  \begin{itemize}
  \item{{\em Similar amounts of vacuum 1 and vacuum 2}} 
  
  We may take, say,
    that the amount of plasma that has come into the big vacuum 2 regions
    is proportional to 1/2*$\Omega_b$.
  \item{{\em $\Delta V = 3\,MeV$ }} 
  
  We assume that the extra energy released
    is due to the potential difference $\Delta V =3\,MeV$ achieved by the passage
    of the nuclei from the vacuum 1 to the vacuum 2.
  \item{{\em Important time interval not expanded away}} 
  
  We assume that the Hubble expansion in say  the last 1/3 of the
    age of the universe was of the order of unity only. So we take it, that
    only the extra energy from the 1/3 of the age of the Universe is
    included.
  \item{{\em $T^4$ behaviour of Energy}} 
  
  We take it, that to get a given
    relative energy increase in the CMB energy density we need a temperature
    change given by the fourth root.
       \end{itemize}

  The extra potential energy achieved per nucleon passing into vacuum 2 is clearly
  \begin{eqnarray}
    \frac{\Delta V}{m_N} &=& \frac{3\,MeV}{1\, GeV}\\
    &=& 0.003 = 0.3 \%.
    \end{eqnarray}
    So comparing this extra energy density to the photon energy density rather than the baryon energy density (using that only half the baryons are now in vacuum 2), we obtain the ratio of the extra energy density to the photon energy density to be 
   \begin{eqnarray} 
     0.003*906*1/2 &\simeq&3*1/2\\
    &=& 1.5.
\end{eqnarray}
   However only the extra energy from the last 1/3 of the age of the Universe is included and thus this ratio is reduced to
    \begin{eqnarray}
     1/3 *1.5&=& 0.5.
     \end{eqnarray}
    This relative shift in $ \Omega_{\gamma} $ of 0.5 gives a shift $\Delta T$ in the CMB temperature T satisfying
    \begin{eqnarray}
    \sqrt[4]{1+0.5}T
    &=& T+\Delta T
    \end{eqnarray}
    which gives 
    \begin{eqnarray}
     \Delta T &\approx & 0.12 T = 0.12 *2.725\, K\\ 
     &=& 0.33\,K
    \end{eqnarray}
    in surprisingly good agreement with equation \ref{deltaT}!


\subsection{Problem of Electric Shielding of Big Bubbles}
\label{s5p2}
Our estimate above takes it as possible and actually occurring that
nuclei from the vacuum 1 penetrate into phase 2, but now there is
the problem, that we argue that a region of phase 2 tends to be surrounded
by an electric field pushing positive charges away. This is because electrons, which are  
supposed not to interact so strongly with the domain wall as the nuclei
or nucleons, would spill over and in the neighbourhood of the vacuum 2
region have a density corresponding to the inside the vacuum 2 region with its
higher density of nuclei due to the lower potential for nuclei.
We estimate that the coat of electrons around a vacuum 2 region will lead
to a potential difference of the order of $\Delta V = 3\,MeV \sim 3*10^6$ volt. This would seemingly keep most cosmic nuclei - except cosmic rays - away, and
they would not be able to penetrate into the vacuum 2 region and
release the energy by passing the domain wall border. So at first it looks
as though the effect, which we used above for explaining the Hubble-Lemaitre-constant
tension, would not function because of this barrier to the penetration of vacuum 2.

However, the full naive stopping by the potential is presumably not 
realistic.
In fact we think that if a nucleus reaches just a little way into the coat of electrons, where the potential begins to rise, it will have a chance to meet
some of the very hot electrons, which actually are the ones causing the potential rise across the coat. 
Then, if so, there is a chance for such an electron hitting the nucleus and transferring enough energy to the nucleus that it manages to pass into the vacuum 2.

Let us make an estimate of what the chance is for such a process:

If a charged nucleus or just a proton has an energy of the order of
$b*\Delta  V$ it can come into the outermost $b$ part of the coat region,
i.e.  penetrate a distance $bd_C$ if the coat thickness is $d_C$. At this distance, due to the Boltzmann factor, the
density of electrons is suppressed by a factor
crudely of the order $\exp(- \frac{d_c -bd_c}{d_c}) = \exp(-1+b)$.    
A small but still of order unity fraction of the electrons are still so energy-rich that they could bring a nucleus to take
up its energy. Then, if not stopped on the way towards the wall, the nucleus will pass the barrier and get caught by the vacuum 2 potential.
The chance that a nucleus passes into vacuum 2 is of course reduced essentially by some small ''exponential" factor, but still not suppressed by any very big factor. If the nucleus should be hit by another electron on the way it might even be helped on its way towards the wall. 

So we would conclude that, in spite of the potential from the coat of electrons pushing the
nuclei away, there will still be a somewhat suppressed probability of penetration into the vacuum 2. It just means that the capture of nuclei will proceed more slowly.
This slowness will presumably be compensated by the density of nuclei
waiting, so to speak, to cross the domain wall and the nuclei will just pile
up depending on how much the crossing is damped. This delaying
of each separate particle will not influence greatly the total passage rate
in the long run.

\subsection{Problem of Fluctuations in CMB}

In our model for the resolution of the Hubble constant tension problem, about 50\% extra energy is dumped into the microwave background radiation in the last 1/3 of the lifetime of the universe. However the fluctuations in the temperature of the CMB as a function of direction is of the order of $10^{-5}$, which requires that the energy is dumped very smoothly. If the heat is spread out sufficiently before getting into the CMB, then a priori
it could smear out the irregularities so much as to give less than the
$10^{-5}$ relative fluctuation as a function of direction.

 When the nuclei (mostly protons) cross the domain walls and are caught by the vacuum 2 regions the heat energy is a priori released at the walls. However we must expect that the heat spreads into both the vacuum 2 and the vacuum 1 neighbouring the walls. At first it seems obvious that it will at least go into
  the vacuum 2 region, because that is where the hot nuclei themselves
  go at first. But they will heat up electrons some of which come into the coat of electrons on the vacuum 1 side of the wall.
  So very much of the heat finds its way to the vacuum 1 also and, with
  sufficient effective time to spread out before the slow process of getting the heat taken up by the CMB, we can hope for enough suppression of the
  fluctuations with direction.

\section{Could the Stopping Length of the Pearl be longer?}
\label{stopping}

We now consider how our dark matter pearls could account for the signal observed by the DAMA-LIBRA experiment 1400 m underground. In our model there is a tension between a pearl reaching down to such a depth and also reproducing the counting rate of events at the DAMA experiment \cite{Bled23}. We shall consider the stopping length of the pearl in the earth above the experiment in this section and the DAMA counting rate in section \ref{solution}.

When the stopping length is estimated using the standard drag force formula and a pearl mass consistent with the DAMA counting rate, the stopping length comes out too small. Here we shall investigate whether the stopping length could actually be longer.

  For relatively large pearls of our dark matter type - bigger than
  say the atoms - we have a very general argument for the drag force $F_D$
  acting on a pearl moving with velocity $v$ through a medium of density $\rho_{medium}$ of the usual form:
  \begin{eqnarray}
    F_D &=& C_D \sigma \rho_{medium}v^2,
    \label{drag}
    \end{eqnarray}
   where $\sigma$ is the cross section of the pearl and the drag coefficient $C_D$ is of order unity. So the equation of motion for a pearl of mass M becomes.
    \begin{eqnarray}
    M\dot{v} &=&-C_D\sigma \rho_{medium}v^2. 
    \label{drag2}
    \end{eqnarray}

Taking $C_D = 1$ and integrating the equation of motion (\ref{drag2}), we obtain
\begin{equation}
	\frac{1}{v} =  t\frac{\sigma\rho_{medium}}{M} + \frac{1}{v_{start}} 
\end{equation}
where $v_{start}$ is the initial velocity of the pearl hitting the Earth of order 300 km/s. The stopping length $L$, where the pearl is moving slowly with a velocity $v_{stop}$, is then
\begin{eqnarray}
L = \int vdt &=& \int 
\frac{dt}{t\frac{\sigma\rho_{medium}}{M} + \frac{1}{v_{start}}}	\\
&=& \frac{M}{\sigma\rho_{medium}} \ln\frac{v_{start}}{v_{stop}}.
\label{dragL}
\end{eqnarray}

   The formula (\ref{drag}) is indeed very suggestively correct, assuming
  the constituents of the medium to be smaller than the pearl and the pearl
  {\em effectively impenetrable} by the constituents:
  
  The particles of the medium being passed by the pearl and which lie in its way have to be moved and they even have to move with the velocity
  of the order of that of the pearl $v$. They must crudely end up with
  kinetic energies of the order of the medium constituent mass $m$ times
  the velocity squared, because they run out of the way with a velocity of
  order $v$. The total mass of the medium constituents 
  expelled per time unit is of course
  \begin{eqnarray}
    \hbox{mass expelled } &=& m\rho_{number}v\sigma,
    \end{eqnarray}
  where $\sigma$ is the area of the pearl seen in the direction of motion and $\rho_{number}$ is the constituent number density.

  This mass crudely picks up from the pearl a momentum of the order of the velocity
  $v$ times this mass per time unit. But that is the force
  \begin{eqnarray}
    \hbox{drag force } &=& v \hbox{expelled mass rate}\\
    &=& m\rho_{number}\sigma v^2 = \rho_{medium}\sigma v^2.
  \end{eqnarray}

  In this section \ref{stopping} we like to investigate, if we can really
  order of magnitudewise trust this drag formula with drag-coefficient of
  order unity to calculate the particle stopping length say. Below we shall find that what can make the stopping length longer than by
  using this dragging formula with of order unity coefficient, is if the
  constituents of the fluid or material through which the pearl moves
  {\em can penetrate smoothly into the pearl and pass through it}, so that they might scatter elastically with a very small angle. Very small  angle
  elastic scattering namely only reduces the speed of
  the pearl by an amount proportional  to the elastic scattering
  angle squared. Thus the stopping length can be
  enhanced by the inverse square of such small scattering angles.
  By interference between constituent particles of the medium passing
  at different places in the transverse direction, the scattering angle can potentially become so small as to be only limited by the finite extension in
  the transverse direction of the pearl and the uncertainty principle.

  But first in section \ref{small} below we shall see that just thinking
  of small particles does not really help modifying the drag formula with a
  coefficient of order unity.

  \subsection{Drag formula for even small particles}
\label{small}
  Here we show that if the cross section $\sigma$ is so small that there
  are long distances between collisions of the pearl on an electron or a
  nucleus, then we also crudely get the drag force formula with the drag
  coefficient of order unity provided the elastic scattering happens with
  an angle of order unity (in radians):

  Denoting in fact the number density of the relevant medium particles 
  by $\rho_{number}$ the distance between collisions becomes
  \begin{eqnarray}
    \hbox{mean free path} &\approx& \frac{1}{\rho_{number}\sigma}.
\end{eqnarray}
The relative momentum decrease per collision is given in order of magnitude by
  \begin{eqnarray}
   \frac{m_{constituent}v}{Mv}
    &=& \frac{m_{constituent}}{M}.
    \end{eqnarray}
    Hence we obtain 
    \begin{eqnarray}
    -M\dot{v}&=&\frac{dx}{dt}* \frac{m_{constituent}}{M}*Mv*
    \sigma*\rho_{number}\\
    &=& \rho*v^2*\sigma,
    \end{eqnarray}
    meaning 
    \begin{eqnarray}
    F_D &\approx& - \rho*v^2*\sigma,
    \end{eqnarray}
  and this is just the drag equation order-of-magnitude-wise with
  $C_D \approx 1$.


  Making the velocity smaller usually makes the Reynolds number smaller
  and thus makes the drag force increase for smaller velocities.
  In our present way of thinking that should probably be
  interpreted to mean that in a viscous fluid the various constituent
  particles, electrons and nuclei, hang together. Thus they effectively
  form objects that could even be bigger than the pearl itself  and
  thereby increase the cross section for hitting such clumps. Note that in
  the argument presented above we could have used the clumps as the
  constituents instead of the electrons and nuclei themselves.

  From the Reynolds number story the cross section for very slow
  particles tends to be effectively bigger and that will make the
  penetration depth  shorter.

  \subsection{Can partly Elastic Clumps help to make a Longer Penetration Depth? }
  \label{interaction}
  But Fermi statistics, can it change the situation?

  If the electron say can only be excited inside the atom by
  an appropriately high enough energy, it looks at first as if the effective cross section could possibly be suppressed. However really it means that 
  the electrons are attached to the atom, and that the atom will just be moved
  as a whole (being a clump in the notation above) and one rather risks the
  effective stopping force to increase.

  But if there are many electrons pushing on the same atom, and they
  push in a random way, because the pearl seeks to push some one way and some
  the opposite way, the net push on the whole atom becomes much less.
  So in this way the pushing of the whole atom could be much smaller than
  if the whole atom had to move away for the pearl.
  Of course it must then mean that the pearl passes through the atom, because
  otherwise it would be pulled away and we would have the argument above.

  Let us attempt to formulate this mechanism convincingly relative to
  the a priori general discussion:

  Remember that in the general discussion of scattering on abstract
  clumps or constituents we had to make the assumption of an order
  of magnitude unity scattering angle, because a very small angle 
  {\em elastic} scattering does not remove momentum in the
  drag direction significantly. So if a clump  is getting a push
  which is an average - averaged in momentum - of many, say $n$, small
  constituents like electrons and nuclei, and these smaller constituents
  only get velocities of order $v$ (the velocity of the pearl) then the
  clump will get a velocity of order $v/\sqrt{n}$ because of the random
  directions
  - at least in transverse directions - and thus the scattering angles
  for the clump become only of the order $1/\sqrt{n}$. It follows that the
  momentum delivery in the longitudinal direction would be down
  compared to the angle of order unity case by a factor $n$.
  (There would though be a problem if the clump gets excited, because then the
  scattering of the clump on the pearl becomes inelastic and the
  scattering brakes the pearl velocity significantly even if the clump
  moves exactly longitudinally after the collision.)
  
    This was a priori a classical description of what could be a mechanism
    for making a larger penetration length for small velocities when
    the constituents of the clump hang together. At high velocity
    the clump is just totally destroyed.

    \subsection{Looking at it as a Smooth Potential}
\label{las}
    In quantum mechanics when we have a clump consisting of a large number $n$ of constituents all interacting with a pearl, because of the uncertainty principle we must think of it as a swarm of constituents with strongly overlapping wave functions. The clump thus becomes a very smooth construction, in the sense that the probability density of the constituents is smoothly spread out over a relatively big region. Now when the pearl approaches the clump, it `` sees'' an extended object of this smooth
    nature and the scattered wave will inherit a smooth form at least in
    the transverse direction. We can think of it as the transverse form of
    the wave function for the pearl achieved by the scattering inheriting an
    approximate translation symmetry. This means that the scattering
    angle {\em cannot} be large/of order unity. So in the elastic case
    we cannot get a big stopping power but only a small stopping power.

    If the wave functions for the constituents in the clump are
    also smooth in the longitudinal direction then even the transfer of
    longitudinal momentum gets suppressed.

    This consideration of the wave function as giving a smooth
    material is only valid, when the pearl is so slow that its pushing on
    the constituents is not able to lift the separate constituents
    up into excited states. So this suppression of
    scattering at large angles and with inelastic excitations
    is a low velocity phenomenon.

    \subsection{Assuming Smooth Pearls}
\label{spp}
    What we hope for is, that we can consider our dark matter pearls
    to act on say a nucleon with a smooth potential, rather than a
    sharp wall. Since the constituents of the pearl are indeed quantum
    fluctuating and even more the constituents of the atoms and molecules,
    there is indeed hope that effectively we have very smooth objects.
    In fact the distances between neighbouring nuclei in our pearl-material
    are about 500 times smaller than in matter under ordinary pressure,
    while we estimated that in ordinary matter the wave functions for the
    nuclei have a size 45 times smaller than the atomic size. So in ordinary
    matter the fluctuating nucleus will fluctuate over several nuclei in the pearl. This should effectively make the pearl a very smooth object. 
    In this case the uncertainty principle suggests the quantum mechanical scattering by our pearls with radius $R$ will provide a transverse momentum of order 1/R.
 
     A nucleus with velocity $v$ has of course the momentum $vm_{nucleus}$,
    and thus if it gets a transverse momentum of the order $1/R$ by
    passing our pearl, it gets a scattering angle of the order
    \begin{eqnarray}
      \hbox{angle } &\approx&\frac{1/R}{vm_{nucleus}}\\
      &=& \frac{1}{Rvm_{nucleus}}
    \end{eqnarray}

    The effect of only small angle scattering is that the stopping power
    is reduced by a factor $(1-\cos(\hbox{angle})) \simeq(\hbox{angle})^2/4$. Hence the drag force $F_D$ is corrected to
    \begin{eqnarray}
    	F_{corrected} &=& \frac{C_D}{4}\sigma\rho v^2*\hbox{angle}^2\\
    	&\simeq& \frac{\sigma\rho v^2}{(Rvm_{nucleus})^2}\\
    	&=&\frac{\sigma\rho}{R^2m_{nucleus}^2}\label{corrected}
    \end{eqnarray}
    provided $R*v*m_{nucleus} >1$. The drag force remains uncorrected if $R*v*m_{nucleus}$ <1. We note that $F_{corrected}$ is independent of the velocity $v$ and hence gives a constant acceleration.

    To get an idea about the orders of magnitude take a nucleus like Na of
    atomic weight 23 and the velocity at the impact of the dark matter
    to earth 300 km/s and use $GeV^{-1} = 0.2*10^{-15}m$. Then
    $v m_{nucleus}= 10^{-3}c *23/(0.2 10^{-15}m) = 10^{14}m^{-1}$.
    So in this example we get small angle scattering provided $R>10^{-14}m$.

    \subsection{Imagine Potential in the Pearls}

    Since we are in the present article deciding to believe the smooth
    potential picture for the interaction of nuclei when they pass our
    dark matter pearls, we shall here put forward our picture of how the pearl
    provides an effective potential for a normal electrically charged nucleus.

    For the purpose of bringing up such a picture we first remark, that while
    the electrons in our dark matter pearls are highly degenerate, the nuclei
    are still rather forming a crystal or a glass, meaning that the motion
    of a nucleon does not cause any severe Fermi statistics problem. So
    where the nucleons are in equilibrium the electric field and the field
    from the potential describing the effect of the vacuum-border would have
    to at least approximately compensate each other locally. We are here
    having in mind that the domain wall is most likely a somewhat extended structure even in the
    normal to its direction. But whatever, we expect the
    electric potential in the stabilized dark matter pearl to  compensate for the potential from the vacuum for average
    nuclei. So in this crudest
    approximation an ``average nucleus'' passing through the dark matter pearl
    will feel no potential except from the electrons, which because of their
    degeneracy are a bit displaced. In fact we would estimate that a layer or cloud of electrons would have been pushed out around the genuine bubble by the effect of the Fermi statistics.
    The whole pearl should be essentially
    neutral and we therefore imagine a relatively thin double layer of
    opposite potential around the pearl. In addition some local irregularities will survive due to the nuclei say in the glass or crystal.

    But such rudimentary potentials will give exceedingly small scattering
    compared to a solid impenetrable pearl, unless the pearl is so thick
    that a nucleus will be met inside the pearl with probability of order
    unity.

    In materials with ordinary pressure and density say 3000 $kg/m^3 = 1.8 *10^{29}GeV/m^3$ there are
    $ 1.8 *10^{29} \hbox{nucleons}/m^3$. Since each nucleon
    delivers a cross section of the order of $\pi *10^{-30}m^2$ we need a
    thickness of the order of $1/(1.8*10^{29}m^{-3}*3.14*10^{-30}m^2$
    = $1/(0.6m^{-1})=1.7m$ for such a collision. 
    
    Compared to our ``ordinary pressure matter density'' $3000kg/m^3$ the  pearl density $\rho_B=5*10^{11}kg/m^3$ we use (\ref{rhoB})
    is $1.7*10^8$ times larger in volume density meaning that the density of
    nuclei say per area is $(1.7*10^8)^{2/3} = 0.31*10^6$ times larger. So 
    the distance a nucleon can on the average penetrate through the inside
    pearl material before hitting a nucleus is $1.7 m/(0.31*10^6)$ = $5*10^{-6}m$. This is larger than the pearl radius $10^{-11}m$ we end up with, and
    the chance of hitting a nucleus is reduced from unity by about a factor $5*10^5$.
    
    This means that when the suppression of the effective cross section by the
    angle squared $(\hbox{angle})^2$ becomes smaller than $2*10^{-6}$, then the hitting of
    a single nucleus in the pearl will take over and we should be back to the
    drag-formula but now multiplied by the probability of hitting $2*10^{-6}$.
    This means that we should not trust the ``smooth'' pearl
    correction to the drag force when it becomes more than a factor $5*10^5$ in the
    stopping length say. But since at least for the drag force the stopping
    length and the radius are proportional, we can crudely say, that the fitted
    radii for the ''smooth" and ''drag f." entries in Table \ref{t} of section
    \ref{table} should not deviate by more than
    $5*10^5$. Luckily for the reliability of our ``smooth'' pearl calculation,
    the deviation is only that the radius needed using the pure drag force
    is $5*10^{-7}m$ being $5*10^4$ (< $5*10^5$) times bigger than the radius $10^{-11}m$  for the ''smooth" one, including the correction with the angle squared.

    \subsection{A Little Problem; The Coat of Electrons}

    But there is the problem with this assumption of a sufficiently smooth
    potential in the pearl felt by a nucleus: 
    In section \ref{ecoat} we estimate  that at the surface of the pearl
    there is a $10^{-12}m$ thick coat of electrons, which
    provides an electric field so strong that the passage through it costs
    of the order of $\Delta V=$ 3 MeV for a proton. The nuclei with galactic
    speeds of the order of 300km/s, have a priori no chance to pass this
    coat. So one would at first think that all protons would have to be expelled
    away from the passing pearl and thus we would be back to the drag formula
    with coefficient unity; the story of the protons etc. passing through the
    pearl seems not to work with this coat taken into account.

    However, the coat is heavily filled with electrons, and we must imagine the
    proton passing through it being screened by electrons around it. If the
    screening length is small compared to the thickness of the coat,
    the $10^{-12}m$, then the force on the screened proton will be much less
    than on the free proton. This could give the proton or nucleus a
    chance to come through into the vacuum 2 region even with the small
    galactic velocity and justify the use of the ''smooth form" of the drag force.
    
    \subsection{Penetration Depth}
    
    Let us first consider the penetration depth calculated from the drag formula (\ref{dragL}) in order to estimate the pearl radius R:
    \begin{equation}
    L = \frac{M}{\sigma\rho_{medium}} \ln\frac{v_{start}}{v_{stop}}.
    \end{equation}
   We take the initial velocity of the dark matter pearl as it impacts the Earth to be $v_{start} =$ 300 km/s $=10^{-3}c$. The ''stopping" velocity $v_{stop}$ is a bit more arbitrary, but the calculation is somewhat insensitive to its value as it occurs inside a logarithmic factor. We take it to be the terminal velocity at which the gravity from the Earth keeps the dark matter particle moving estimated using the drag force to be of order 25 m/s. Thus the logarithmic factor becomes in order of magnitude 
   \begin{equation}
   ln\frac{v_{start}}{v_{stop}} = ln\frac{3*10^5m/s}{25 m/s}= 9.4
   \end{equation}
   
   Using the pearl density $\rho_B = 5*10^{11}kg/m^3$ evaluated in section \ref{homolumo}, we obtain
   \begin{equation}
   \frac{\sigma}{M}= \frac{\pi R^2}{4\pi R^3\rho_B/3} = \frac{3/4}{5 *10^{11}kg/m^3}\frac{1}{R} \label{inversedarkness}
   \end{equation}
and hence
\begin{eqnarray}
	 L &=&
	9.4\frac{M}{\sigma\rho}\\
	&=&
	9.4*5*10^{11}kg/m^3*4/3*\frac{R}{3000kg/m^3}\\
	&=&
	2.1*10^9R.\label{prop}
	\end{eqnarray}
 We now require $L \simeq 1$ km the depth of the DAMA-LIBRA experiment, giving the radius of the pearl to be
 \begin{eqnarray}  
  R&=& \frac{10^3m}{2.1*10^9} \\
 &=& 5*10^{-7}m.\label{Rpd}
 \end{eqnarray}

    \subsection{Stopping with ``smooth form'' of Drag Force}
\label{smooth}
  Let us now estimate the pearl radius R by calculating the penetration depth from the corrected ''smooth form" of the drag force (\ref{corrected})
  \begin{eqnarray}
  F_{corrected} =\frac{\sigma\rho}{R^2m_{nucleus}^2}
  \end{eqnarray}
For definiteness we consider a Na nucleus with
\begin{equation} 
	m_{nucleus} = 23*(0.2*10^{-15}m^{-1}) = 10^{17} m^{-1}
\end{equation}
  This corrected form gives the pearl a constant acceleration $-a$, where
  \begin{eqnarray}
  	a&=& \frac{\sigma\rho}{MR^2m_{nucleus}^2}\\
  	&=& \frac{3/4*3000kg/m^3}{5*10^{11}kg/m^3*10^{34}m^{-2}*R^3}\\
  	&=& \frac{4.5*10^{-43}m^2}{R^3}
  \end{eqnarray}
  The stopping length is taken when the velocity is zero and we neglect gravity:
  \begin{eqnarray}
  	L&= &\frac{v_{start}^2}{2a}\\
  	&=&\frac{10^{-6}}{9*10^{-43}m^2}R^3,	\\
  	&\approx& \frac{R^3}{10^{-36}m^2}
  \end{eqnarray}
which we require to be $L \simeq 10^3 m$ and hence the pearl radius R to be given by
  \begin{eqnarray}
  	R^3 &=& 10^{-33}m^3\\
  	\hbox{or} \quad R &=& 10^{-11}m.\label{Rps}
  \end{eqnarray}

  \section{Homolumo Gap and Density of Pearl Material}
 \label{homolumo}
 
 Our dark matter pearls consist of very highly compressed ordinary matter inside a new vacuum bubble. A homolumo energy gap $E_H$ is supposed to appear between the filled and empty electron states. In reference \cite{theline} we have estimated the order of magnitude of this energy gap in terms of the Fermi energy or momentum $E_f =cp_f$, using a dimensional argument or more precisely the Thomas-Fermi approximation:
 \begin{eqnarray}
 	E_H &=& \sqrt{2} \left ( \frac{\alpha}{c}\right )^{3/2}E_f. 
 \end{eqnarray}
 Here $\alpha$ is the fine structure constant which in our somewhat special dimensional argument was taken to be of dimension velocity, so that it is $\alpha/c$ that is the usual fine structure constant 1/137.036...
 
 We identify $E_H$ with the 3.5 keV energy of the X-ray line
 mysteriously observed by satellites from clusters of galaxies, Andromeda
 and the Milky Way Center \cite{Bulbul,Boyarsky,Boyarsky2} and attributed to dark matter. In this way we estimate the Fermi momentum to be $\sim 3$ Mev. The electron number density in the interior of our pearls is therefore
 \begin{eqnarray}
 	n_e &=& \frac{1}{3\pi^2}p_f^3\\
 	&=& \frac{1}{3\pi^2}(3\, MeV)^3\\
 	&=& 1.4*10^{38}\, m^{-3}.
 \end{eqnarray}
 We take it that there are 2 nucleons per electron and then we obtain the mass density $\rho_B$ of the pearl matter to be:
 \begin{eqnarray}
 	\rho_B &=& 2m_N*n_e\\
 	&=& 5*10^{11}kg/m^3, \label{rhoB}
 \end{eqnarray}
where $m_N = 1.7 * 10^{-27}$ kg is the mass of a nucleon.
 
 \section{DAMA-like Experiments}
 \label{DAMAlike}
 
 The underground experiments, ANAIS \cite{ANAIS}, DAMA-LIBRA \cite{DAMA2}, and  a COSINE-100 \cite{cos100oct22} looking for dark matter are very similar:
 \begin{itemize}
 	\item
 	They use NaI(Tl = Thallium),
 	\item
 	They look for seasonal variation,
 	\item
 	Depths are 4200 m.w.e. for DAMA-LIBRA, 2450 m.w.e. for ANAIS and 1600 m.w.e. for 
 	COSINE-100
 \end{itemize}

 The annual modulation amplitudes from the various experiments \cite{cos100Yu} are given in Table \ref{modulation}.

 
 \begin{table} [h]
  \begin{tabular}{|c|c|c|}
    \hline
Counts/kg/keV/day&

1–6 keV&

2–6 keV\\
\hline

This work Cosine-100&

- 0.0441 $\pm$ 0.0057&

- 0.0456 $\pm$ 0.0056\\
\hline

DAMA/LIBRA&

0.0105 $\pm$ 0.0011&

0.0095 $\pm$ 0.0008\\
\hline

COSINE-100 &

0.0067 $\pm$ 0.0042&

0.0050 $\pm$ 0.0047\\
\hline

ANAIS-112 &

- 0.0034 $\pm$ 0.0042&

0.0003 $\pm$ 0.0037\\
\hline
\end{tabular}
\caption{  The amplitudes of the annual modulation fits using the DAMA-like method
  to the COSINE-100 3 years data (this work) are compared with results from
  DAMA/LIBRA, COSINE-100, and ANAIS in both 1–6 keV and 2–6 keV
  regions.} \label{modulation}
\end{table}
ANAIS sees no evidence for modulation in their data, which is incompatible with the DAMA-LIBRA result at $3\sigma$. 
 
\begin{figure}
\includegraphics[scale=0.14]{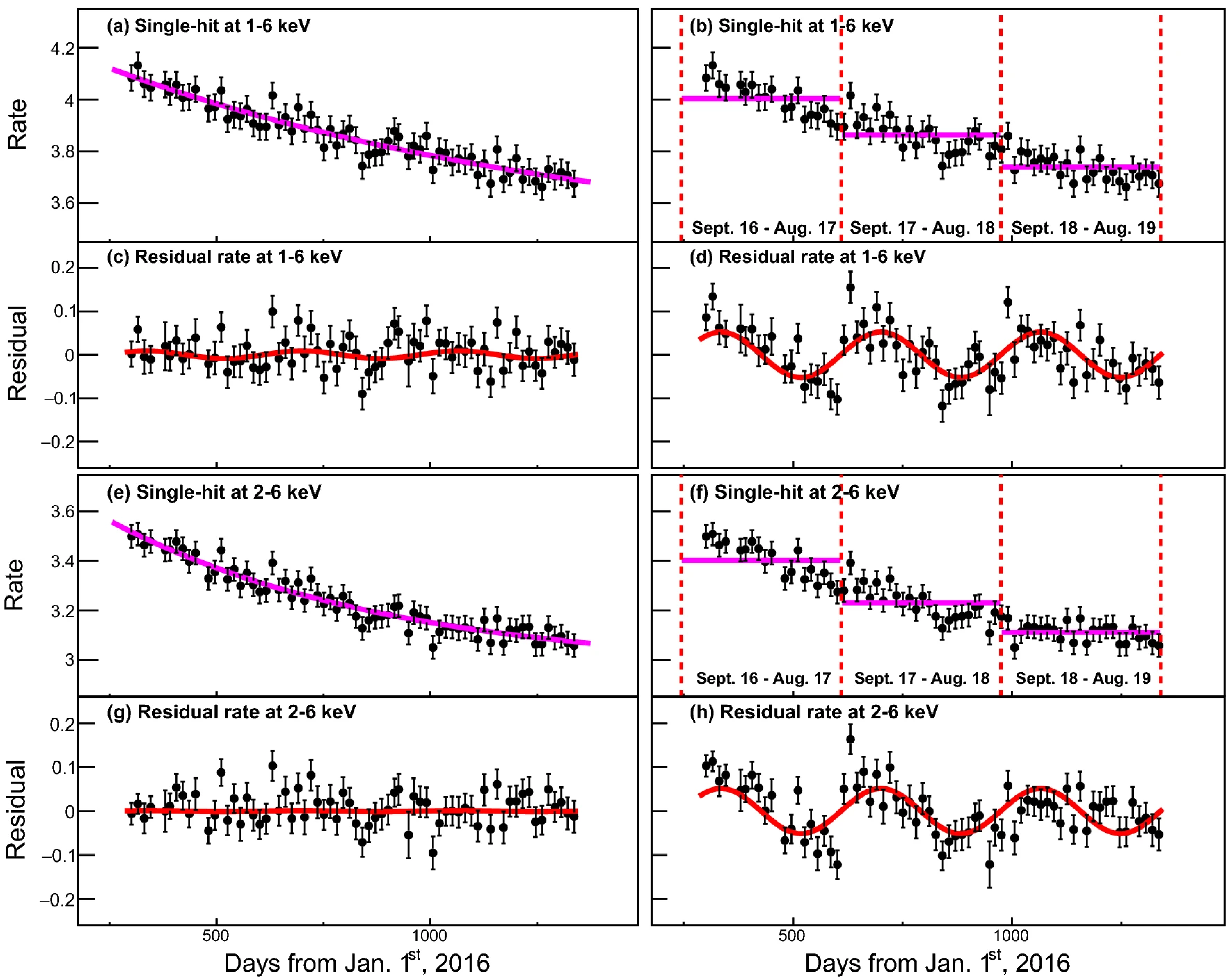}
\caption{Single-hit event rates in the unit of counts/keV/kg/day as a function of time. The top four panels present time-dependent event rates and the residual rates in the single-hit 1–6 keV regions with 15 days bin. Here, the event rates are averaged for the five crystals with weights from uncertainties in each 15-day bin. Purple solid lines present background modelling with the single exponential (a) and the yearly averaged DAMA-like method (b). Residual spectra for the single exponential model (c) and the DAMA-like model (d) are fitted with the sinusoidal function (red solid lines). Same for 2–6 keV in the bottom four panels. Strong annual modulations are observed using the DAMA-like method while the result using the single-exponential models are consistent with no observed modulation.}
\label{Damalike}
\end{figure}

The COSINE-100 group searched for an annual modulation in their data with the phase fixed at the DAMA value of 152.5 days. They subtracted their calculated time dependent background and obtained a positive amplitude of $0.0067 \pm 0.0042$ cpd/kg/keV, which is to be compared with the DAMA-LIBRA amplitude of $0.0105 \pm 0.0011$ cpd/kg/keV. The large error on the COSINE-100 amplitude means that it is consistent with both the DAMA-LIBRA result and no seasonal variation at all. But then they analysed their own COSINE-100 data in the ''same way as DAMA" \cite{cos100Yu} by subtracting a constant background taken to be the average over one year. In this way they generated what they consider to be a spurious seasonal variation as shown in Figure \ref{Damalike}. In fact they found a modulation amplitude of the opposite phase $-0.0441 \pm 0.0057$ cpd/kg/keV ( i.e. with more events in winter than in summer). Interestingly the WIMP model could not obtain such a result, but in principle it is possible with our dark matter pearls \cite{Bled23}. 

%

  The solution to the disagreement between the experiments in our model is that our dark matter pearls interact so strongly that they get {\em slowed down} in the earth shielding, and have stopping lengths such that they first essentially stop or move slowly at a depth of the order of that of the DAMA-LIBRA experiment. The main signal for the observation of the pearls is {\em radiation} of electrons or photons with the preferred  energy of 3.5 keV as discussed in the previous section. But the dark matter pearls pass through the ANAIS and COSINE-100 experiments so fast that there is not sufficient time for them to give an observable signal.

  \section{DAMA-LIBRA Event Rate}
  \label{solution}

 Let us first note the estimated value of the $\frac{\sigma}{M}$ parameter or
 "inverse darkness" of our pearls. It is given by the formula (\ref{inversedarkness})
   \begin{equation}
  	\frac{\sigma}{M}= \frac{3/4}{5 *10^{11}kg/m^3}\frac{1}{R} 
  \end{equation}
  Using the radius $R = 5*10^{-7}m$ estimated from the simple drag formula gives
   \begin{equation}
  	\frac{\sigma}{M}|_{drag f.}= 3*10^{-6}m^2/kg = 3*10^{-5}cm^2/g.
  	\label{dragdarkness}
  \end{equation}
  But when we believe in the more penetrable pearl and use the radius $R =10^{-11}m$ estimated from the "smooth form" of the drag force suppressed by a factor (angle)$^2$ we obtain
   \begin{equation}
  	\frac{\sigma}{M}|_{smooth}= 0.15m^2/kg = 1.5cm^2/g.
  \end{equation}
   
  Now let us consider the DAMA-LIBRA event rate:
  
   The density of dark matter in the region of the solar system is
     \begin{eqnarray}
       D_{sun} &=& 0.3\, GeV/cm^3
       \end{eqnarray}
       The rate of impact energy on the Earth from dark matter is thus 
    \begin{eqnarray}   
       ``Rate'' &=& v*D_{sun}\\
       &=& 300 km/s * 0.3 *1.79*10^{-27}kg/(10^{-6}m^3)\\
       &=& 1.5*10^{-16}kg/m^2/s\\
       &=&1.3*10^{-11}kg/m^2/day   
     \end{eqnarray}
 It follows that:
 \begin{equation}
\hbox{Number density of pearls falling on Earth} = \frac{1}{M} 1.3*10^{-11}kg/m^2/day
\end{equation}  
 Then giving each kg NaI say $10^{-2}m^2$, we obtain the number of pearls passing through 1kg of NaI per day in the DAMA detector as
 \begin{equation}
 	 (100kg/m^2)^{-1}\frac{1}{M}1.3*10^{-11}kg/m^2/day =
 	 \frac{1}{M}1.3*10^{-13}/day
 \end{equation}


   The connection of this number  $\frac{1}{M}1.3*10^{-13}/day$ to the observations of DAMA requires a
   slight interpretation and a hypothesis:

   What DAMA gives us is the number of counts per day per kg NaI per keV being
   \begin{eqnarray}
     \hbox{``DAMA rate''} &=&0.0103 cpd/kg/keV 
     \end{eqnarray}
   The first thing to do is to sum over the of order 3 keV that are active to get
   \begin{eqnarray}
     3keV *\hbox{``DAMA rate''} &=&0.0103 cpd/kg/keV*3keV\\
     &=& 0.03 cpd/kg
     \end{eqnarray}
   Next we must take into account that what DAMA here observed was the
   {\em oscillating part } of the dark matter flux. But, since the
   velocity of the Earth around the sun is only about 1/10 of the velocity
   of the sun relative to the average dark matter, only about 1/10 of the
   flux of dark matter comes in the oscillating mode this way. Thus the
   presumed full
   intensity should be
   \begin{eqnarray}
     \hbox{``Full flux ''} &=& 10 *0.03 cpd/kg\\
     &=& 0.3cpd/kg.
     \end{eqnarray}

   Now there are a priori several possibilities in our model:
   \begin{itemize}
   \item{{\em Constant Interaction.}} It can be that effectively our pearl interacts constantly when it moves sufficiently slowly along through
     the matter/the apparatus of DAMA; in that case we should then simply
     identify
     \begin{eqnarray}
       0.3 cpd/kg &=& \frac{1}{M}*1.3* 10^{-13}cpd\\
       \Rightarrow M&=& 4.3 *10^{-13}kg
     \end{eqnarray}
       Using the pearl density $\rho_B = 5*10^{11}kg/m^3$, we obtain the pearl radius
       \begin{eqnarray}
       R &=& \sqrt[3]{M/(\rho_B*4\pi/3)}\\
       &=& \sqrt[3]{4.3 *10^{-13}kg/(5*10^{11}kg/m^3*4\pi/3)}\\
       &=&6*10^{-9}m\label{Rra}
     \end{eqnarray}
   \item{\em Few shots. } If the pearls only send essentially one
     shot of say electrons of 3.5 keV, when they have stopped it will
     only be rather few of the pearls that accidentally stop in the scintillator material of DAMA compared to the number going through the apparatus without being seen.

     In such a case the ``measurement'' of the here written $0.3 cdp/kg$
     means that there could be of the order of $1km/0.1m=10^4$ times
     more unnoticed passages of the apparatus which we took to be 0.1 m thick.
     That would mean that we rather have
     \begin{eqnarray}
       10^4*0.3cpd/kg&=& \frac{1}{M}*1.3*10^{-13}cpd\\
       \Rightarrow M&=& 4.3*10^{-17}kg\\
       \hbox{So } \quad R &=& \sqrt[3]{4.3*10^{-17}kg/(5*10^{11}kg/m^3*4\pi/3}\\
       &=& 0.27*10^{-9}m\label{Rro}
       \end{eqnarray}
   \end{itemize}

   \section{Review of Mass and Size of Pearl}
   \label{table}
   
   Here we like to collect the estimates of the mass $M$ of our pearls and
   the corresponding radius $R$ - when we use the by the homolumo gap
   estimate gotten density of the pearl-matter $\rho_B=5* 10^{11}kg/m^3$ -
   Since we should also like to present the corresponding cubic roots $S^{1/3}$
   of the tension $S$ of the wall around the pearl, we use the fact that the pressure $P$ exerted by the wall is balanced by the electron degeneracy pressure to give 
   \begin{equation}
   	2S/R = P = \frac{p_f^4}{12\pi^2}
   \end{equation}
   Hence using $p_f = 4$ MeV we have
   \begin{eqnarray}
     S& =&R*1.08 MeV^4\\   
     &=& R*\frac{1.08}{0.198*10^{-12}}MeV^{3}/m\\
     &=&R*5.45*10^{12} MeV^{3}/m\\
     \hbox{So }\ \quad \sqrt[3]{S}&=&\sqrt[3]{R}*1.76 *10^4MeV/m^{1/3}\label{l176}
   \end{eqnarray}

  Let us also note that with the density of $\rho_B = 5*10^{11}kg/m^3$
  the mass is given as
  \begin{eqnarray}
    M&=& \frac{4\pi}{3}\rho_B*R^3\\
    &=&2.1*10^{12}kg/m^3 *R^3. \label{massp} 
    \end{eqnarray}
  \begin{table} 
   \begin{tabular}{|c|c|c|c|c|c|c|c|}
     \hline
     No. &From &Assumption&Mass $M$& Radius $R$& Tension& $S^{1/3}$&Ref.\\
     \hline
     1.&Rate&All seen&$4.3*10^{-13}kg$&$6*10^{-9}m$&$3.3*10^4MeV^3$
     &$32 MeV$&(\ref{Rra})\\
     \hline
     2.&Rate&One shot&$4.3*10^{-17}kg$&$2.7*10^{-10}m$&$1.5*10^3MeV^3$&12MeV
     &(\ref{Rro})\\
     \hline
     3.&Stop length& smooth&$2.1*10^{-21}kg$ &
     $10^{-11}m$&$
     55\, MeV^3 $&
     4 MeV
     &(\ref{Rps})\\
     \hline
     4.&Stop length &drag f.& $2.6*10^{-7}kg$&$5*10^{-7}m$&$4.4*10^{6}MeV^3$&
     $120 MeV$&(\ref{Rpd})\\
     \hline
     5.&Cosmology& One layer& $4.5*10^{-13}kg$&$6*10^{-9}m$&$3.3*10^4MeV^3$&
     $32 MeV$&(\ref{S3cm})\\
     \hline
     6.&Cosmology& 100 layers&$4.5*10^{-19}kg$ &$6*10^{-11}m$&$330\, MeV^3$&$7 MeV$&\\
     \hline
   \end{tabular}
   \caption{\label{t} Table of various estimates of radius and mass under the
     assumption that the pearl density is $\rho_B = 5*10^{11}kg/m^3$ (\ref{rhoB}).}
\end{table}

   \subsection{Comments on the Table of Mass and Size Estimates}

   First it should be said that the purpose of Table \ref{t} is to compare
   mass or size estimates from various observations and theoretical
   assumptions named in the columns number 2 and 3. The following
   four columns actually give just one number, because the four numbers
   in the columns numbered  4 to 8 are just mathematically related numbers
   for each row. The density of the interior
   of our pearls is taken (see section \ref{homolumo})  as $5 *10^{11}kg/m^3$. The tension $S$ ensures that the pearl can balance the electron degeneracy pressure from the internal matter with the radius in question.
   The first column just enumerates the rows, and the final eighth column refers to the formula in our
   article, in which the result of the relevant calculation is given for the row in question.

   The second column refers to the main measurement or basic information and the third column refers to any assumption used
   in the lines in question:
   \begin{itemize}
   \item{{\em Rate.}} Rate means that in this line we fit to the observed rate
     of events in the DAMA experiment.
     \begin{itemize}
     \item{{\em All seen.}} With the calculation under the name ``All seen''
       we assume that our pearl radiates so strongly - say electrons
       of 3.5 keV - that whenever a pearl passes the experimental  apparatus
       of DAMA, it gets noticed by giving a count. For this to be true a rather
       high intensity of radiation of electrons from the pearl is needed.
     \item{{\em One shot.}} In this calculation we take it that the pearl
       probably gets effectively stopped in a range of depths of order
       1 km, and that it delivers all of its electron radiation at the stopping
       point. This emission is conceived of as just one event
       by DAMA. 

       It could of course well be that it emitted many electrons and that
       thus this ``One shot'' estimate would underestimate the number of
       events to be seen by DAMA, or it could be that not all the pearls got excited and it would be an over-estimate.
     \end{itemize}
   \item{{\em Stop length}.} The ``observation'' underlying this `` Stop length ''
     is that, in order to achieve the somewhat mysterious fact that no 
     underground experiment looking for dark matter other than DAMA sees
     any dark matter, we assume that the dark matter shall just have
     a penetration depth equal to the 1400 m depth of DAMA. 
     So the first assumption is that the penetration depth is of order
     of 1 km.

     Next we present the prediction according to how we imagine the potential that the pearl provides for the particles, nuclei or electrons, hitting it:
     \begin{itemize}
     \item{{\em smooth.}} The possibility ``smooth'' means that in fact
       the pearls offer a very smooth
       and not very high potential to the particles in the shielding. So the constituents of the shielding
       can pass only mildly disturbed across the pearl and thus typically
       these constituents scatter with a small scattering angle. We
       take it that this angle is determined by the cross section of the pearl,
       since smaller angles than those given this way can hardly
       be realized.
    
     \item{{\em drag f.}} This notation ``drag f.'' is meant as
       shorthand for use of the drag force formula with the drag coefficient
       being of order unity. This drag force formula with unity coefficient
       comes as a result of taking the pearl to be quite impenetrable for the
       constituents of the shielding. The important assumption separating this
       drag force hypothesis from the ``smooth'' case is that the scattering
       angle is assumed almost all the time to be large or rather of order
       unity (in radian units).
       \end{itemize}
     
     \item{{\em Cosmology}}
     \begin{itemize}
     	\item{{\em One layer.}}
     	In section \ref{cosmology} we estimate the highest value of the surface tension a single domain wall having a dimension of the visible Universe length scale could have ($S=(32MeV)^3$), 
        without its energy dominating out the
     	total energy density integrated over the visible Universe.
     	So if it had a higher tension than that, $S> (32MeV)^3$, it would have totally spoiled our models of cosmology. We use this upper limit in the table.
     	\item{\em {100 layers.}}
     	In this case we instead use the assumption that layers of domain walls
     	cover the whole Universe with a distance between the  layers such
     	that there are 100 layers across the visible Universe. That is to say
     	there should be approximately parallel layers with a distance between them of the	order of 1/100 of the size of the visible Universe, 13 milliard light years. I.e. there would be about 100 Mly (mega-light-years) between the walls. Surrounding voids by domain walls would correspond to something like this and
     	would thus require this $1/\sqrt[3]{100}$ of the 32 MeV upper limit
     	for the third root of the tension.

     \end{itemize}
     \end{itemize}

   \subsection{Conclusions on Tension from Table \ref{t}}
   \label{SRM}
  
   The most trustable entries in Table \ref{t} are the items in
   line 2. and 3. namely, ``Rate One shot'' and ``Stop length smooth''
   which agree within expected accuracy. We take the average value of $\sqrt[3]{S}=8$ MeV as the best fit to the domain wall tension. The corresponding radius (\ref{l176}) and mass (\ref{massp}) are R = $10^{-10}$ m and M = $2*10^{-18}$ kg = $10^{9}$ GeV.
   
   The best fit domain wall tension is therefore
   \begin{eqnarray}
     S &=& ( 8 MeV)^3\\
     &=& (8 MeV *1.78*10^{-30}kg/MeV)*(8 MeV/(0.198*10^{-12}MeVm)^2\\
     &=& 2.3*10^{-2}kg/m^2\\
     &=& 23.1 g/m^2. \label{human}
     \end{eqnarray}
   
   Using that 1 kg corresponds to $(3*10^8m/s)^2kg = 9*10^{16}J$
   we get that this domain wall tension is
   \begin{eqnarray}
     S &=& 2.3*10^{-2}kg/m^2*9*10^{16}J/kg\\
     &=& 2.1*10^{15}N/m
     \end{eqnarray}
   The mass per area (\ref{human}) is of very human scale
   but the tension of $2.1*10^{15}N/m$ is enormous.

  The domain wall presumably has a thickness of the order of $1/(8 MeV)$, which is of atomic nucleus size in order of magnitude.
  The sound velocity along the domain wall is of course essentially light
   velocity, when the wall stands still.
   Just to press an extremely thin needle of atomic radius $10^{-10}m$ and
   perimeter around $10^{-9}m$ into it would need a force $2.1*10^6 N$, the weight of 210 ton. This will also be the force needed to hammer the pearl
   out of shape, unless the material inside is even harder.

   \subsection{The Electron Coat of Vacuum 2 Regions}
   \label{ecoat}
   The potential $\Delta V \simeq 3$ MeV for nucleons between the two types of vacuum is not supposed to be felt directly by the electrons.  We therefore expect that whether we have an electron degenerate dark matter pearl or a large hot region of vacuum 2, some of the electrons will end up outside the vacuum 2 region although kept near the pearl or bubble because of the  electric attraction to the protons in the nuclei in the vacuum 2 region. 
  We must expect that it is dominantly the pressure of the electron
   gas, degenerate or just hot, which keeps the bubble spanned out. The nuclei
   which do not provide much of the pressure are essentially free to adjust
   their positions so that the electric force on them and the force from the
   phase of vacuum (thought here a bit smeared out) cancel each other on the
   average nucleus.  Thus the electric potential in the interior
   (vacuum 2 side) will be just $\Delta V= 3$ MeV setting the vacuum 1 potential
   to $0$. Now there will be a coat of electrons, the thickness of which,
   say $d_C$,  we want to estimate. For continuity reasons the density of the
   electrons in the coat will still be order of magnitudewise the same as
   inside the vacuum 2 region. Now the electron number density inside the pearl is
   \begin{eqnarray}
     \rho_{number\; e}&=& 5*10^{11}kg/m^31/2 *6*10^{26}GeV/kg \\
     &=& 1.5 *10^{38}electrons/m^3.
     \end{eqnarray}
  So the charge density $\rho_{el}$ in a region with the density of
  electrons like in the interior of the pearl, but with the nuclei removed
  (by the vacuum potential), is  
     \begin{eqnarray}
    \rho_{el} &\sim & -1.5 *10^{38}electrons/m^3 *1.6 *10^{-19}C/electron\\
     &=& -2.4*10^{19}C/m^3.
     \end{eqnarray}
  
  The Maxwell equation for these electrons is
   \begin{eqnarray}
     div \vec{E}&=& \rho_{el}/\epsilon_0,
     \end{eqnarray}
     which for flat layering becomes
     \begin{eqnarray}
     \frac{\partial E}{\partial x_{transverse}} &=&
     \frac{\rho_{el}}{\epsilon_0}\\
     \hbox{or } \qquad -\frac{\partial \partial V_{el}}{\partial^2x_{transverse}}&=& \frac{\rho_{el}}{\epsilon_0}.
     \end{eqnarray}
     Order of magnitudewise we therefore have
     \begin{eqnarray}
     (\Delta V)_{in \; volt} /d^2_C&=& -\rho_{el}/\epsilon_0   
     \end{eqnarray}
     and hence
     \begin{eqnarray}
     d^2_C &\approx & \frac{\epsilon_0(\Delta V)_{in \; :volt}}{-\rho_{el}} \label{ffd}\\
     &=& \frac{(8.854*10^{-12}F m^{-1})*3*10^6 Volt }{2.4 *10^{19}C/m^3}\\
     &=& 1.1*10^{-24}m^2.
     \end{eqnarray}
     Thus the thickness of the coat of electrons around a pearl is 
     \begin{eqnarray}
     d_C&\approx& 10^{-12}m 
   \end{eqnarray}



   For the big hot bubbles out in the Universe, for which we think of
   having densities of matter like the average density in the whole
   Universe, we have say one atom per cubic meter. This means the density of electrons for the big extended bubbles like the big voids is
   \begin{eqnarray}
     \rho_{el\; big \; bubble} &=& 1.6*10^{-19}C/m^3.
     \end{eqnarray}
     Using (\ref{ffd}) we then have 
     \begin{eqnarray}
     d^2_C &=& \frac{(8.854*10^{-12}Fm^{-1})*3*10^6V}{1.6*10^{-19}C/m^3}\\
            &=& 1.7 *10^{14}m^2.
      \end{eqnarray}      
     So the thickness of the coat of electrons around a big bubble is
      \begin{eqnarray}      
            d_C &=&1.3*10^{7}m\\
            &=& 1/30\,  \hbox{light-second}.
     \end{eqnarray}

   \subsection{Interaction in (Dwarf) Galaxies}

   The interaction between dark matter and dark matter has been observed
   by various studies of the influence on visible objects, especially
   Correa \cite{CAC} has studied numerically dwarf galaxies and
   also collected data from larger objects as shown 
   in Figure \ref{CC}. 

   \begin{figure}
     \label{CC}
     \includegraphics{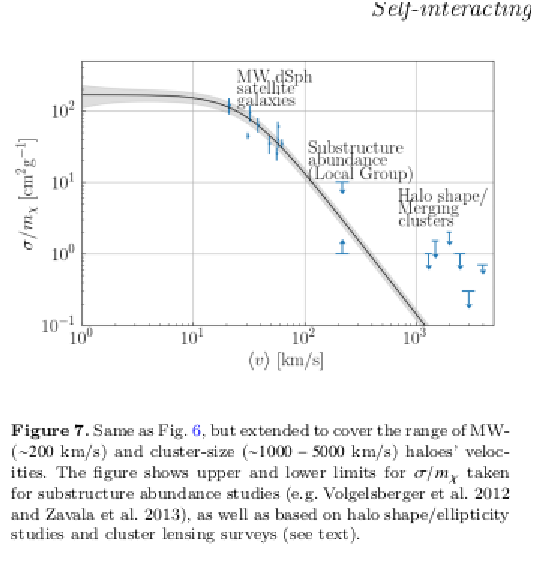}
     \caption{
       \label{CC}
       The collection of indirect observations of the ratio $\frac{\sigma}{M}$ for the self-interaction of dark matter
       versus the velocity of the dark matter particles. Very crudely
       the majority of the data-points fall on a skew line, which in the double logarithmic plot has the slope -2 corresponding to the formula
       for this region
         $\frac{\sigma}{M} \approx \frac{1.2*10^{10}m^4/(kg s^2)}{v^2}$.  
     }
   \end{figure}

   The maximal $\frac{\sigma}{M}$ obtained by Correa is rather large compared to the value $\frac{\sigma}{M}|_{drag f.}$ (\ref{dragdarkness}) needed to get the penetration depth of order 1 km according to the simple drag formula. So we have previously
   studied \cite{Corfu21} the possibility that out in interstellar space the dark matter pearls are dirty in the sense of very likely carrying a dust clump with them, that can easily be appreciably larger
   although still light compared to the dark matter bubble of new vacuum.
   Also remember that the plots in Figure \ref{CC} are for dark matter interacting with itself, while the interaction relevant for whether or not the dark matter reaches the DAMA experiment is its interaction with ordinary matter. 
   We must therefore take into account both the following points
  
   \begin{itemize}
   \item Dark matter self interaction can be different from
     it interaction with ordinary matter (baryons, atoms).
   \item The dark matter in outer space is likely to be dirty, while the
     dark matter may have been cleaned in the atmosphere, when it hits
     the Earth.
   \end{itemize}

   We shall therefore postpone the comparison of the interaction of dark matter in the (dwarf) galaxies and the interaction with the shielding above the underground experiments till
   a later article.

   Let us though remark, that the variation in the best measured region of
   velocities of the inverse darkness $\frac{\sigma}{M}$ with velocity in Figure \ref{CC} is $\propto \frac{1}{v^2}$. This is the same dependence we used in the ``smooth form" of the drag force, namely that the effective cross section goes down with $v$ because the scattering
   angle could goes as $\propto 1/v$ and the stopping effect then goes as the square of this angle.
   Also the value of $\frac{\sigma}{M}|_{smooth} = 1.5 cm^2/g$ obtained using the  ``smooth form" of the drag force is of the same order as the $v= 300$ km/s point on Figure \ref{CC}.
   
  \section{Low Domain Wall Tension}
\label{cosmology}

  The tension or in $c=1$ notation also the energy per area $S$ of a domain
  wall between different phases of the vacuum has dimensionality $GeV^3$. If
  an energy scale of the order of the scales at which we look for
  `` new physics'' is used the energy density gets so high that domain walls
  of cosmological scales of extension would be so heavy as to be totally
  excluded by the already known energy density, the critical density
  \begin{eqnarray}
    \rho_c &=& \frac{3H^2}{8\pi G}= 1.8788*10^{-26}h^2kg/m^{-3}.
  \end{eqnarray}
   We shall use the CMB value (\ref{CMBh}) from the Planck Collaboration \cite{CMB} for the Hubble constant
   $h=H/(100km/Mpc) = 0.674$, which gives
  \begin{eqnarray}
    \rho_c &=& 8.5*10^{-27}kg/m^3.
    \end{eqnarray}
  
 Consider a domain wall having a dimension of the visible Universe length scale:
  \begin{eqnarray}
    R_{visible} &=& 13 *10^9 \; \hbox{light} \; \hbox{years}\\
    &=& 1.3*10^{10}ly *9.5*10^{15}m/ly = 1.2 *10^{26}m
    \end{eqnarray}
   giving an area 
    \begin{eqnarray}
    Area &\sim & 10^{52}m^2.
    \end{eqnarray}
    Now the total visible critical energy is
  \begin{eqnarray}
    \hbox{visible (critical) energy} &\sim & (1.2 *10^{26}m)^3
    *8.5*10^{-27}kg/m^3\\
    &=& 1.5*10^{52}kg.
  \end{eqnarray}
  So the maximum energy per area $S_{max}$ allowed for such a domain wall is
  \begin{eqnarray}
   S_{max} &\sim& \frac{1.5*10^{52}kg}{10^{52}m^2}
    \sim  1.5 kg/m^2\\
    &=& 1.5*2.2*10^4 MeV^3 = 3.3*10^4 MeV^3\label{Scm}
    \end{eqnarray}
 Hence the maximal energy scale for such aa domain wall is 
\begin{eqnarray}
	 S^{1/3}_{max} &\sim& 32\, MeV \label{S3cm}
	\end{eqnarray}
 
Now our best fit to the tension in the domain wall around a pearl (\ref{l176}) is given in section \ref{SRM} as
      \begin{eqnarray}
        S^{1/3}_{Pearl}&=& 8\, MeV
              \end{eqnarray}
      This is the energy scale of {\em pion or hadron physics}
      indicating that the {\em two phases of the vacuum we propose are to
        be distinguished by some pion or hadron physics!}
      \section{Why the Big Regions of Vacuum 2 are Extremely Hot}
\label{hot}
From the perspective of general ''new physics" a domain wall tension $S^{1/3}$ of the order of a relatively few MeV is extremely low, while for cosmological purposes it is a highly significant tension.
Let us now look at what temperature $T$ a large region of vacuum 2 with size $R$ could have with a matter density $\rho_{number}$ measured for instance in atomic units (or GeV) per $m^3$.

 Measuring the molar density $n/V$ in numbers of atoms per $m^3$
called here $\rho_{number}$, we have
\begin{eqnarray}
	n/V &=& \rho_{number}/N_A,
\end{eqnarray}
where $N_A$ is the Avogadro constant. The gas constant is $R = N_A *k$, where $k$ is the Boltzmann constant and so the gas equation $PV = nRT$  may be written in the form
\begin{equation}
	P =\rho_{number}kT
	\end{equation}
	
The pressure needed to keep a spherical bubble of radius $R$ spanned out
is
\begin{eqnarray}
	P&=& 2S/R
	\end{eqnarray}
	\hbox{so we need } 
	\begin{eqnarray}
	2S/R &=& \rho_{number} kT.\label{pressure}
\end{eqnarray}

We now consider three representative values of the tension $S$, which a cosmological domain wall could have, and using (\ref{pressure}) calculate the corresponding temperatures $T$ of 100 Mpc size vacuum 2 bubbles which we show in Table \ref{Temperature}:
 \begin{itemize}
	\item{{\em $S = (3\,MeV)^3$.}}
	\begin{eqnarray}
		S &=& (3\,MeV)^3\\
		&=& 3^3\frac{1.6*10^{-13}\,J}{(0.198*10^{-12}\,m)^2}\\
		&=& 1.1* 10^{14}\,J/m^2\\
		&=& 1.1*10^{14}\,Pa\,m\\
		&=& 3^3\frac{1.78*10^{-30}\,kg}{(0.198*10^{-12}\,m)^2}\\
		&=& 1.2*10^{-3}\,kg/m^2.
	\end{eqnarray} 
	\item {\em $S = (360\,MeV)^3$}
	\begin{eqnarray}
		S &=& (360\,MeV)^3\\
		&=& 1.9*10^{20}\,Pa\,m\\
		&=& 2.1*10^{3}kg/m^2. 
	\end{eqnarray}
		\item {\em $S = (8\,MeV)^3$}
	\begin{eqnarray}
		S &=& (8\,MeV)^3\\
		&=& 2.1*10^{15}\,Pa\,m\\
		&=& 2.3*10^{-2}kg/m^2. 
	\end{eqnarray}
	We mostly believe the $S=(8\,MeV)^3$ value.
	
  \end{itemize}

  \begin{table}
      \begin{tabular}{|c|c|c|c|c|c|}
        \hline
          R&$\rho_{number}$ & For $S=(3MeV)^3$&For $S=(8MeV))^3$& For $S=(360 MeV)^3$&\\
          &&=$1.1*10^{14}$Pa m &= $2.1*10^{15}$ Pa m& =$1.92*10^{20}$ Pa m&\\
          &&gives&gives& gives&\\
          && $kT_{S^{1/3}=3MeV}$ &$kT_{S^{1/3}=8 MeV}$& $kT_{S^{1/3}=360 MeV}$ &\\
          \hline

          100 Mpc&$1a.u./m^3$&$7.1*10^{-11}J$&$1.4 *10^{-9}J$&$1.2*10^{-4}J$&\\
          = $3.09*10^{24}m$&&=$ 5.2*10^{12}K$ &=$10^{14}K $&
          =$8.9*10^{18}K$ &\\
          &&=$4.4*10^2 MeV$&=$8.7*10^3 MeV$&=$7.7 *10^8$MeV&\\
          \hline
          100 Mpc&$1a.u./cm^3$&$7.1*10^{-17}J$&$1.4*10^{-15}J$&$1.2*10^{-10}J$&\\
          = $3.09*10^{24}m$&&=$ 5.2*10^6K$ &=$10^{8}K $&
          =$8.9*10^{12}K$ &\\
          &&=$4.4*10^{-4} MeV$&=$8.7*10^{-3} MeV$&=$7.7 *10^2$MeV&\\
          \hline
      \end{tabular}
      \caption{Temperatures of 100 Mpc size vacuum 2 bubbles with matter density equal to that for the whole Universe $\rho_{number} = 1\, a.u./m^3$ and for a much higher density $\rho_{number} = 1\, a.u./cm^3$.} \label{Temperature}
      \end{table}

      If you want to have a minimum of one bubble of vacuum 2 in each void
      of 100 Mpc size then, since the total density in the whole Universe is
      of order $1 a.u./m^3$ ordinary matter, you cannot get the inside density
      $\rho_{number}$ above one, unless you let the radius go down with the
      cubic root of the number $\rho_{number}$ compared to the void size of
      100 Mpc. 
      Let us, for example, take the tension in the wall to be    $S=(360\,MeV)^3$ so that we see from Table \ref{Temperature} it would get the absurd temperature $7.8*10^8MeV$. Suppose now that we should like to bring this temperature down by a factor $10^8$ to a few MeV. Then we must take $R\rho_{number} = 10^8$ (\ref{pressure}) and bring the radius of the bubble down by a factor of $10^4$ to 1/100 of a Mpc and just have one bubble in the whole void. So we better   
      hope for a low tension $S$ in the domain walls, if we shall have big
      bubbles.

      But in any  case we see that there are troubles in getting the bubbles
      colder, if they shall not be so few and small that they only cover a tiny bit of the Universe.

      One way out of this problem could be to have percolating vacua. That
      is to say we could have some huge regions of say vacuum 2
      covering several large parts of different voids so as to make the connected
      vacuum 2 region extend infinitely or very much longer than only a
      few voids. One can in fact have what is called percolation, that both
      phases vacuum 1 and vacuum 2 occur in infinitely extended regions. If so,
      then it is possible that the domain wall between them would have 
      - at least over most of the surface - eigen-curvatures of opposite sign, so that
      the average curvature at many points of the wall could be a sum of a
      negative and a positive curvature. Then it is possible, if the wall has
      put itself in such a configuration, that the pressure needed to keep the
      wall in place could be
      much smaller than the typical curvature of the wall, by compensation.
      This would of course mean that the wall had negative Gauss-curvature and
      thus locally have the character of a saddle point surface.

      By avoiding spherical vacuum 2 regions and instead have regions with a
      negative Gauss curvature border, the need for a terribly strong
      pressure could be reduced and somewhat colder regions could be
      possible.
      
      \subsection{Significance of Hot Bubbles}

      The significance of these very very hot bubbles of vacuum 2 is that
      they very easily lose their heat, if they have a chance to do so.
      But to see how this can come about we must first think about how they
      are surrounded by a layer of electrons:

      We suppose that it is mainly the nucleons that feel a strong potential
      stopping them from escaping the bubble. This is because we suspect that
      the formation of the domain walls is related to the strong interactions.
      The estimated order of magnitude of the tension indicates strong
      interactions as the suggested mechanism; also electromagnetic
      interactions are too well understood that any as yet unknown new phase
      could hide there. So the electrons, having by Coulomb interactions
      achieved an exceedingly high temperature of the order of $\Delta V$
      say, would leak out of the bubble and spread - thus cooling off the
      bubble at the surface - were it not that they would form an electric
      field near the surface of the bubble, that would drive them back.
      In the long run Coulomb fields would ensure that no appreciable amount of electrons can leave the bubble including its nearest surroundings without taking protons with them. Now to stop electrons of temperature
      $\Delta V$ requires an electric potential 
      of essentially same size as $\Delta V$. This is a very large
      electric potential for external protons to meet, when they are about to
      penetrate into the bubble. So all positively charged particles from
      outside the bubble except exceedingly hot ones (cosmic rays) will
      meet a potential seeking to push them out when attempting to 
come into the big bubble. But we argued in section \ref{s5p2} that
by the help of interaction with the electrons, which are very hot,
we expect that with a time-delay some protons and nuclei will
come into vacuum 2 anyway. But those which don't will come into the outer part of the coat of electrons
      around the bubble and have a chance of close contact - via the Coulomb
      potential say - with this coat of electrons. Thereby they have a chance
      to cool the bubble, which would then shrink by such cooling if
      significant
      amounts of matter come into the coat of electrons around the bubble.

      The chance of protons meeting the electrons in the coat and thus cooling the large bubble
      is higher in the galactic rich regions than in the voids, simply because
      there is more matter there even in the space around the galaxies or
      between stars inside the galaxies. Thus the possibility for a bubble to
      retain its high temperature is bigger in the voids than in the galaxy
      rich regions. So we expect that the bubbles of vacuum 2 outside the voids will tend to cool down and freeze to death, in the sense that they
      will contract and presumably end up as ordinary dark matter, based on having a clump of degenerate electron matter (metal in the atomic sense) as we assume the usual dark matter is.
      
\section{Tension S; what is the Physics Distinguishing the Two Vacua?}
\label{behind}

In this section we consider the physics that fixes the energy scale of the surface tension $S$ of the domain wall separating the two vacua.
      The cubic root of the tension $S$ in the wall surrounding the pearls is of the order
      \begin{eqnarray}
        \sqrt[3]{S} &=& 8 MeV, 
      \end{eqnarray}
      which suggests:
      \begin{itemize}
      \item The underlying physics should be {\em Pion or Hadron Physics}: In
        fact we have found that there is
        a possibility for there being a phase transition in the Standard Model vacuum, on the one
        side of which the Nambu Jonalasinio spontaneous breaking is really a
        spontaneous breaking, while in the other vacuum it is rather that the quark masses have just broken this symmetry.
      \item If we speculated that domain walls {\em surrounded the voids}
        seen
        in between the galaxy rich regions with extensions
        of order of 30 to 300 Mpc, then
        the {\em energy density of the  walls could replace the dark
          energy density.}
      \end{itemize}
      
\begin{figure}
      \includegraphics[scale=0.9]{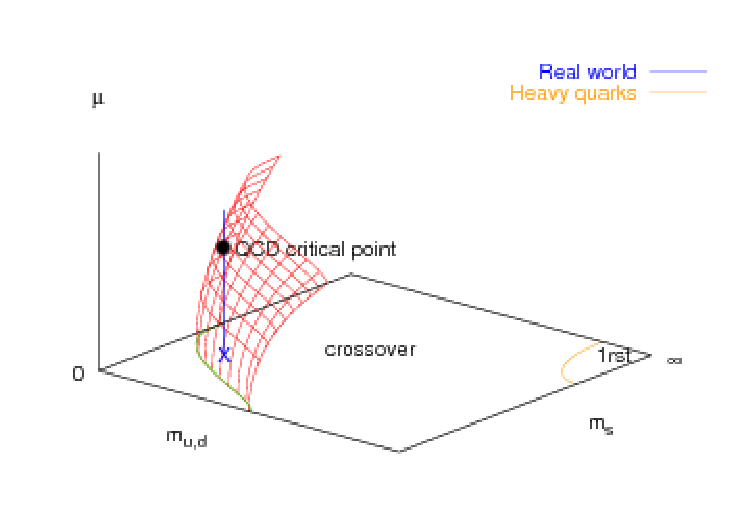}
      \caption{Here is a 3-dimensional phase diagram \cite{Columbiaplot} in which there is imposed a chemical potential $\mu$ for baryon number along the vertical axis, while at the bottom in perspective we see the usual Columbia plot with the common $u$ and $d$ quark mass and the strange quark mass.}
      \label{Forcrand}
      \end{figure}

\begin{figure}
      	\vspace{-3cm}
      	\includegraphics[scale=0.6]{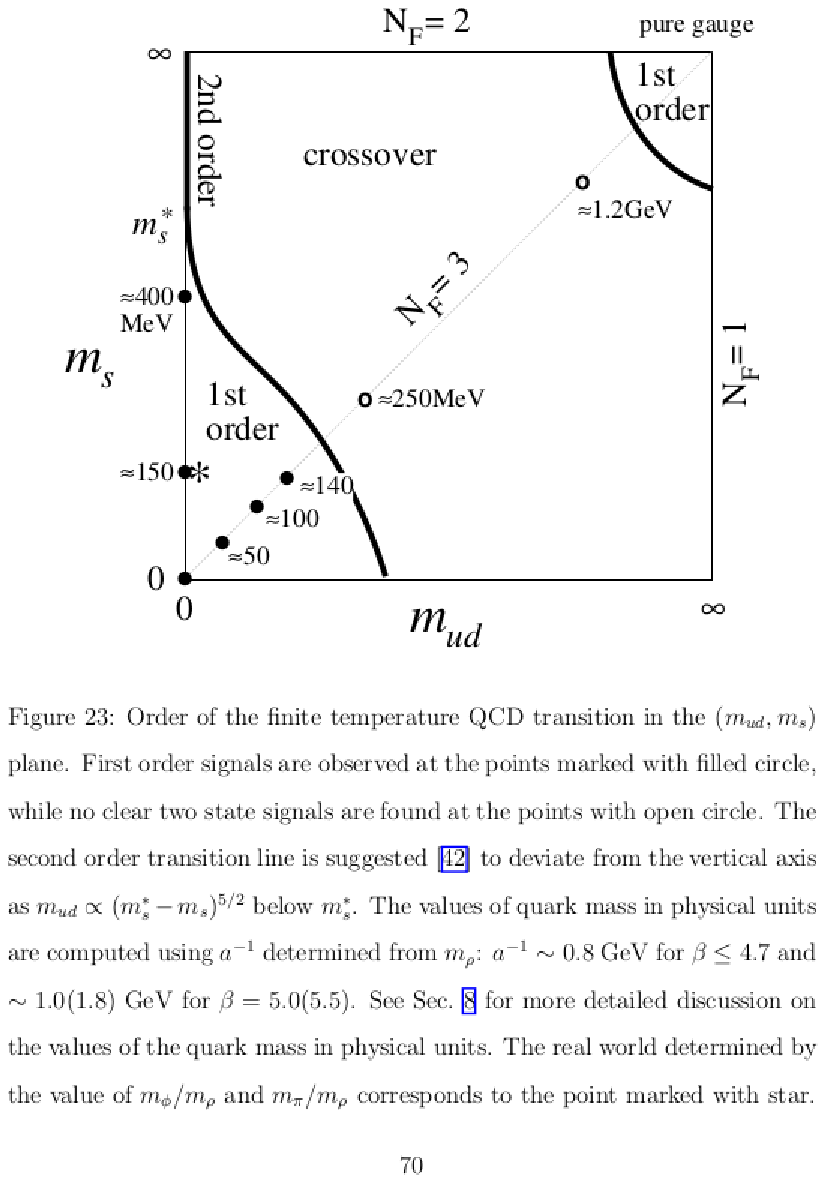}
      	\caption{This plot is taken from reference \cite{Japanese}, which used lattice calculations to find the phase transition temperatures and whether there was a true first order phase transition or just a crossover. The phase transition temperatures are associated to the points in quark mass plane at which the temperature was found by their lattice calculations.}
      	\label{Twasaki}
      \end{figure}

      The QCD phase diagram involving quark masses is the most relevant for us. In Figure \ref{Forcrand} a schematic picture of the QCD phase diagram at low temperature is shown \cite{Columbiaplot}, as a function of the quark masses and the matter density represented by the quark chemical potential $\mu$. 
       The floor in the plot is the zero chemical potential
      surface, corresponding to the phase diagram for the vacuum in the quark mass plane (simplified to just the strange quark mass $m_s$ and the assumed common mass of the two light quarks $m_u=m_d$), known in the literature as the ''Columbia plot". The green line indicates the presence of a second order phase transition curve and the surface swept out by this critical line as $\mu$ increases may intersect the physical quark masses' vertical line and give a high density QCD critical point. The main property investigated in the Columbia plot is how raising the temperature can cause phase transitions. In the lower quark mass corner of the plot there is a spontaneous symmetry breaking of the chiral symmetry and a first order phase transition occurs as the temperature is raised. For larger quark masses there is no true corresponding phase transition, but instead just a crossover. If there is a different behaviour of a phase transition as a
      function of temperature in different regions of quark mass space, it must mean that the regions in quark mass
      space are in different phases. This is illustrated in Figure \ref{Twasaki} from an earlier paper \cite{Japanese}.
      
\begin{figure}
      \includegraphics[scale=0.85]{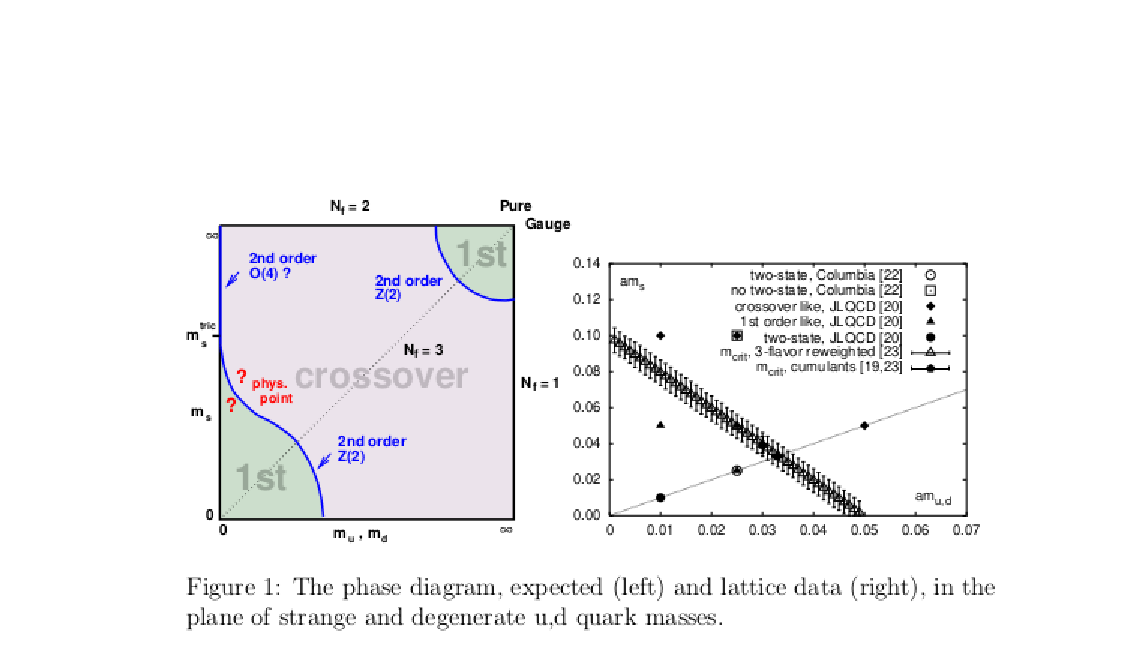}
      \caption{Here Laermann and Philipsen \cite{question} have left question marks to tell that they did not know on what side of the second order phase transition the physical combination of quark masses lies.}
      \label{Laermann}
      \end{figure}

      For us the crux of the matter is that there {\em is} a phase transition
      in the quark mass plot somewhere so close to the actual quark masses,
      that we can have the dream/speculation that Nature - because of the
      ``multiple point criticality principle'' - could have chosen to fine tune the parameters of the Standard Model, especially the quark masses, to just sit on the second order
      phase border. Note that the physical quark masses are on the lower mass side of the second order phase border in Figure \ref{Twasaki} (denoted by the symbol $\ast$) but on the higher quark mass side in Figure \ref{Forcrand}. This ambiguity is explicitly exhibited in Figure \ref{Laermann} of reference \cite{question} by two question marks, one on each side of the second order phase transition curve. This possibility, that
      one does not really know in which phase Nature has put the quark masses
      is very important for our speculations. We namely speculate that the
      Nature-point lies just on the phase transition curve. But if so of course it would not be possible with realistic accuracy to settle in which of the phases the experimental quark mass combination lies. However we must admit that more recent lattice calculations  mostly find the physical quark masses to lie in the crossover region with higher quark masses \cite{Guenther}.

      A phase transition of this type would  involve the Nambu
      Jonalasionio spontaneous breakdown and have an energy scale of the
      order of say the pion mass. We found that our domain wall had a cubic root of its tension in the energy range of about
      8 MeV.

      Remember : We claim a {\em principle} that there shall be {\em many
        vacuum-phases} fine-tuned (as a new physics principle) to have
      the same energy density.

      \section{Conclusion}
      \label{conclusion}
    
      We have put forward elements of our model of dark matter as small bubbles of a new vacuum phase containing highly compressed ordinary matter with particular emphasis on the Hubble constant tension problem:
      
      \begin{itemize}
      \item We have for many years proposed the fine-tuning { \em Multipoint Criticality Principle} (MPP), according to which there should be
        {\em several vacuum phases with the same energy density}.
        
      \item If the energy density per domain wall area - the tension $S$ - is
        as small as suggested
        by our work on our dark matter model, say of order $S \sim (8\,  MeV)^3$,
        then we could have {\em astronomical size regions of new vacuum}, e.g.
        filling out the big voids observed with rather few galaxies in them.
        Or perhaps there could be some sort of percolating space-regions, the two vacua
        percolating each other, but as large as the voids say.
      \item Such astronomically large new vacuum regions are supported by evidence for MPP:
        \begin{itemize}
        \item 4 $\sigma$ evidence for {\em variation of the fine structure
          constant $\alpha$} by of the order of $10^{-5}$ relatively.
          
        \item It could be helpful for the {\em $\theta$ term fine-tuning problem}.
          
        \item Some time ago we used our MPP to {\em  PREdict the Higgs mass} (before the
          Higgs particle was found) to be $135\, GeV \pm 10\, GeV$. Later it was, of course, measured to be $125 GeV$.
        \end{itemize}
        
      \item {\em The degenerate vacuum phases can plausibly alleviate the Hubble constant
        tension problem}. In fact doing so mainly by releasing energy when ordinary
        matter (baryons) cross the domain wall going into the
        ``new phase'', in which the potential energy is lower by $\sim 3\, MeV$.
        Then, by the Suniaejev-Zamalogikov (SZ) effect, this released energy heats up the microwave background
        radiation to the observed 2.725 K. But one should then really use the
        value, hopefully $\simeq$ 2.4 K, had the SZ effect not taken place.
        This helps to allow a higher Hubble constant than expected from the usual calculation using the
        Planck CMB data.
        
        \item Assuming that ordinary matter nuclei can pass through the dark matter 
        bubble we managed to consistently fit the size of the bubble to about
        $R =10^{-10}\, m$, in the sense that this size agrees with both the
        penetration depth being about the $1km$ depth of DAMA-LIBRA
       and the counting rate of particles observed in DAMA-LIBRA. This size corresponds to a
        mass for the bubbles of $M=2*10^{-18}\, kg = 10^9\, GeV$ and a tension for the domain wall of
        $S=(8\, MeV)^3$, when using the density of the interior of the bubble
        $5*10^{11}kg/m^3$ and the X-ray spectral line attributed to dark matter or the homolumo gap being $3.5 keV$.
      \item Led by the order of magnitude of the tension $S \sim (8\,MeV)^3$ in the surface between the two vacua, we speculate that a phase
        transition found by QCD lattice calculations as a function of the
        quark masses could be identified with the phase transition connected
        with our two vacua.
        \end{itemize}
      \section*{Acknowledgement}
      One of us HBN thanks the Niels Bohr Institute for his emeritus status and use of an office. Also CDF would like to thank Glasgow University for his Honorary Senior Research Fellowship.

\end{document}